\PassOptionsToPackage{table}{xcolor}
\pdfoutput=1
\documentclass[sigconf]{acmart}
\AtBeginDocument{%
  }

\acmSubmissionID{2470}

\author{Ruyi Ding}
\orcid{0000-0002-0079-8265} 
\authornote{These authors contributed equally.} 
\affiliation{
  \institution{Louisiana State University}      
  \city{Baton Rouge}
  \country{USA}}
\email{ruyiding@lsu.edu}

\author{Tianhong Xu}
\authornotemark[1]
\affiliation{
  \institution{Northeastern University}      
  \city{Boston}
  \country{USA}}
\email{xu.tianh@northeastern.edu}

\author{Xinyi Shen}
\affiliation{
  \institution{Yale University}      
  \city{New Haven}
  \country{USA}}
\email{xinyi.shen@yale.edu}

\author{Aidong Adam Ding}
\affiliation{
  \institution{Northeastern University}      
  \city{Boston}
  \country{USA}}
\email{a.ding@northeastern.edu}

\author{Yunsi Fei}
\affiliation{
  \institution{Northeastern University}      
  \city{Boston}
  \country{USA}}
\email{y.fei@northeastern.edu}

\usepackage{xspace}
\usepackage{xcolor}
\usepackage{listings}
\usepackage{float}
\usepackage{hyperref}
\usepackage{enumitem}
\usepackage{subcaption} 
\usepackage{dsfont}
\usepackage{tabularx}
\usepackage{multirow}
\usepackage{makecell}
\usepackage{longtable}
\usepackage{pifont}
\newcommand{\cmark}{\ding{51}}%
\newcommand{\xmark}{\ding{55}}%
\setlist[enumerate]{topsep=0pt, partopsep=0pt, itemsep=0pt, parsep=0pt}
\setlist[itemize]{topsep=0pt, partopsep=0pt, itemsep=0pt, parsep=0pt}
\definecolor{BrickRed}{RGB}{203,65,84}
\definecolor{MidnightBlue}{RGB}{25,25,112}
\definecolor{ForestGreen}{RGB}{34,139,34}
\definecolor{YellowGreen}{RGB}{154,205,50}
\definecolor{VioletRed}{RGB}{208,32,144}
\lstdefinestyle{mystyle}{
    basicstyle=\ttfamily\small\color{black!80},
    keywordstyle=\color{blue},
    commentstyle=\color{green!50!black},
    stringstyle=\color{red},
    numbers=left,
    columns=fullflexible,
    keepspaces=true,
    showspaces=false,
    showstringspaces=false,
    breaklines=false,
    frame=tb,
    framerule=0.5pt,
    xleftmargin=2em,
    framexleftmargin=1.5em,
    tabsize=2,
    float=H,
    morecomment=[l]{//}, 
    keywords={func, for, if, return, continue}, 
    escapeinside={/*@}{@*/} 
}

\newcommand{\mymethod}{\textrm{MoEcho}\xspace}

\newif\ifcomment
\newcommand{\rd}[1]{\ifcomment {\color{purple}
\emph{[Ruyi: #1]}} \fi}
\newcommand{\ad}[1]{\ifcomment{\color{green}
\emph{[Adam: #1]}}\fi}
\newcommand{\yf}[1]{\ifcomment{\color{red}
\emph{[Fei: #1]}}\fi}
\newcommand{\tx}[1]{\ifcomment{\color{blue}
\emph{[Tianhong: #1]}}\fi}

\newcommand{\update}[1]{\textcolor{black}{#1}}
\usepackage{tikz}
\newcommand{\tikzcircle}[2][blue,fill=blue]{\tikz[baseline=-0.5ex]\draw[#1,radius=#2] (0,0) circle ;}%

\begin{document}

\title{MoEcho: Exploiting Side-Channel Attacks to Compromise User Privacy in Mixture-of-Experts LLMs}

\begin{abstract}
    The transformer architecture has become a cornerstone of modern AI, fueling remarkable progress across applications in natural language processing, computer vision, and multi-modal learning.
    As these models continue to scale explosively for performance, implementation efficiency remains a critical challenge.
    Mixture-of-Experts (MoE) architectures, selectively activating specialized subnetworks (experts), offer a unique balance between model accuracy and computational cost. 
    However, the adaptive routing in MoE architectures—where input tokens are dynamically directed to specialized experts based on their semantic meaning—inadvertently opens up a new attack surface for privacy breaches. 
    These input-dependent activation patterns leave distinctive temporal and spatial traces in hardware execution, which adversaries could exploit to deduce sensitive user data.
    In this work, we propose \textbf{MoEcho} (MoE-Echo), discovering a side-channel analysis-based attack surface that compromises user privacy on MoE-based systems. 
    Specifically, in MoEcho, we introduce \textbf{four novel architectural side-channels} on different computing platforms, including \textit{Cache Occupancy Channels} and \textit{Pageout+Reload} on CPUs, and  \textit{Performance Counter} and \textit{TLB Evict+Reload} on GPUs, respectively.
    Exploiting these vulnerabilities, we propose \textbf{four attacks} that effectively breach user privacy in large-language models (LLMs) and vision-language models (VLMs) based on MoE architectures: \textit{Prompt Inference Attack}, \textit{Response Reconstruction Attack}, \textit{Visual Inference Attack}, and \textit{Visual Reconstruction Attack}.
    We evaluate MoEcho on four open-source MoE-based models at different scales, with a specific focus on the DeepSeek architecture. 
    Our end-to-end experiments on both CPU- and GPU-deployed MoE models demonstrate a $99.8\%$ success rate in inferring the patient's private inputs in healthcare records and $92.8\%$ in reconstructing LLM responses.
    MoEcho is the first run-time architecture-level security analysis of the popular MoE structure common in modern transformers, highlighting a serious security and privacy threat and calling for effective and timely safeguards when harnessing MoE-based models for developing efficient large-scale AI services.
    
\end{abstract}

\maketitle

\section{Introduction}
The transformer architecture has become the cornerstone of modern AI systems~\cite{vaswani_attention_2017,lin_survey_2022,khan_transformers_2022}, driving breakthroughs in diverse applications such as natural language processing~\cite{wolf_transformers_2020}, computer vision~\cite{wu_visual_2020}, and multi-modality learning~\cite{xu_multimodal_2023}. 
However, as AI models continue to scale rapidly for  performance~\cite{kaplan_scaling_2020}, implementation efficiency has emerged as a critical challenge. 
To address this issue, various techniques have been developed~\cite{tay_efficient_2022}, including Key-Value Caching~\cite{pope_efficiently_2022, ge_model_2024} to eliminate redundant computations, fast attention mechanisms to optimize the attention operations~\cite{dao_flashattention_2022,dao_flashattention-2_2023}, and quantization methods to reduce the precision during model inference~\cite{lin_awq_2024,zhao_atom_2024}. 
Among the innovations, the Mixture-of-Experts (MoE) architecture uniquely balances efficiency and accuracy~\cite{dai_deepseekmoe_2024,jiang_mixtral_2024,shazeer_outrageously_2017,zoph_st-moe_2022,lin_moe-llava_2024} with adaptive computations.

MoE-based transformers are designed to dynamically route input tokens to specialized subnetworks—commonly referred to as \textit{experts}—that perform task-specific computations. 
Each expert is trained to process distinct input features (e.g., syntax, semantics, or domain-specific knowledge). 
A lightweight routing network then selects the most relevant experts for each input token based on an affinity score.
For example, in DeepSeekMoE~\cite{dai_deepseekmoe_2024} for the DeepSeek-V2 Lite model~\cite{deepseek-ai_deepseek-v2_2024}, there are 64 experts available, but each token only activates  6 most suitable ones, reducing computational cost by 85\% during inference while preserving performance.
MoE architectures have been widely adopted in modern large language models, including Gemini-1.5~\cite{gemini2024unlocking}, Mixtral~\cite{jiang_mixtral_2024}, Grok-1~\cite{xai_grok1_2023}, the just-released Llama 4~\cite{meta2025llama} and  DeepSeek-R1~\cite{deepseek-ai_deepseek-r1_2025,deepseek-ai_deepseek_2024,dai_deepseekmoe_2024}.
However, as the expert activation pattern in MoE-based models is highly correlated with users' inputs, it becomes a privacy vulnerability and effective exploitation may lead to user privacy breaches. 
\yf{is this right? what is the baseline for keeping and using 64 experts? is it Lite compared to a full-blown model?} \rd{yes. This is what we use in the experiment}
\yf{what do you mean for expanding the model's capacity here?} \rd{I remove this claim. One of the advantages of MoE is it allow the model using more parameters during training (while keep the same size for inference. But this may cause confusion here.}

\begin{figure}[th]
    \centering
    \includegraphics[width=0.99\linewidth]{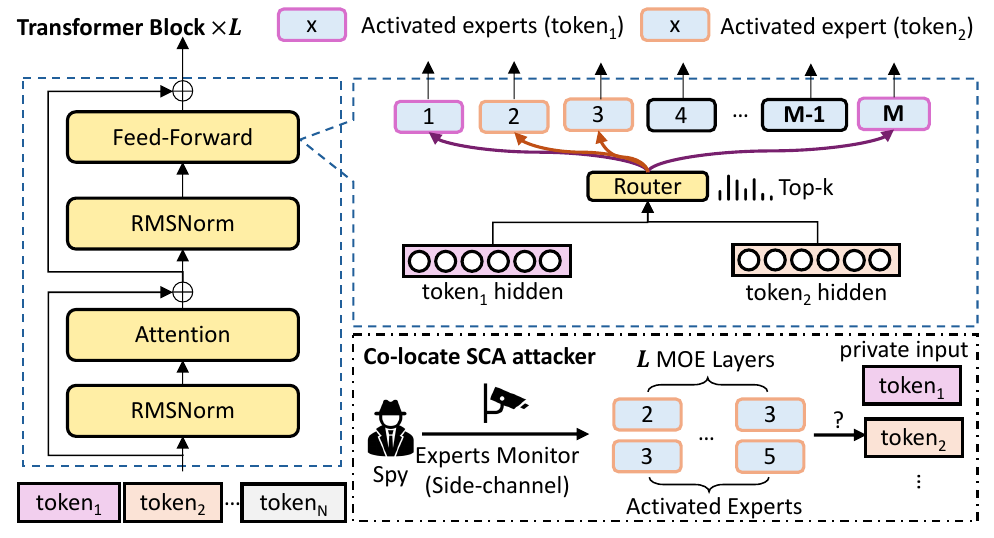}
    \vspace{1mm}
    \caption{\textit{Motivation of \mymethod}. The expert activation pattern of the MoE system is highly dependent on the user inputs. Therefore, \mymethod leverages side-channel analysis to retrieve these patterns and infer the user's private inputs.}
    \Description{This figure is the motivation for the vulnerability of Mixture of Experts}
    \label{fig: moe-motivation}
\end{figure}

In this paper, we discover a new attack surface intrinsic to the MoE architecture, arising from its input-dependent execution patterns.
Unlike dense models with fixed computational paths, where the model is responding uniformly to different inputs, MoE's dynamic routing leaves distinguishable footprints: both temporal (i.e., varying execution time of experts due to different workloads) and spatial (i.e., expert-specific memory access patterns), presenting a new vulnerability.
We propose \textbf{\mymethod} (MoE-Echo), \textbf{Mixture-of-Experts Echoing}, to exploit architectural side-channels that leak these expert execution footprints so as to infer user private information (e.g., the input prompts). Figure~\ref{fig: moe-motivation} illustrates the motivation for \mymethod on a decoder-only LLM, where each input token in the user prompts or the sequentially generated response token will activate various experts, leaving vulnerable footprints.

Nevertheless, input-dependent execution patterns, both temporal and spatial, cannot be easily captured during complex LLM execution by software-based methods, e.g., monitoring the timing differences through controlled prompt injection~\cite{yona_stealing_2024}. 
Considering the ample architectural side-channels on different computing platforms, we resort to side-channel analysis (SCA) to obtain these execution patterns~\cite{yan2020cache, zhang2024invalidate+, jiang2017novel, yarom2014flush+}.
Existing SCAs of ML systems primarily compromise confidentiality, targeting model parameter extraction~\cite{gongye2024side} or structure retrieval~\cite{yan2020cache}. 
Our work focuses on user privacy and investigates various architectural side-channels that can leak MoE execution patterns. 
Specifically, we design four side-channels for two different deployment platforms.
For models deployed on CPUs, we propose \textbf{(a) cache occupancy channels} to explore temporal leakage of experts' executions, and \textbf{(b) Pageout+Reload Attacks} designed to track the memory access patterns of experts.
We also consider MoE on GPU, and propose to utilize the \textbf{(c) GPU Performance Counters} to monitor the expert load with massive parallel computations, and \textbf{(d) TLB Evict + Reload Attacks} to determine the spatial footprints of experts on GPUs.

Based on the side-channels, we propose four novel privacy attack vectors, targeting both MoE-based large language models and multi-modal models. Specifically, our attacks include:
\textbf{(1) Prompt Inference Attack:} This attack aims to recover the user's input prompts along with their associated attributes, effectively revealing sensitive linguistic cues,
\textbf{(2) Response Reconstruction Attack:} By leveraging side-channel information during the subsequent tokens generation, this attack reconstructs the model’s output generated in response to user queries,
\textbf{(3) Visual Attribute Inference Attack:} Focused on the vision modality, this attack deduces key attributes (e.g., facial landmarks) or identities embedded within input images,
\textbf{(4) Visual Reconstruction Attack:} This attack conditionally reconstructs the user’s input images with generative models, providing insights into the visual content processed by the model.

Through these comprehensive attack methodologies, our work exposes critical vulnerabilities inherent in the dynamic routing and execution patterns of MoE architectures, especially on DeepSeekMoE—one of the most widely adopted open-source MoE implementations, highlighting the urgent need for robust security measures in modern AI systems.
The contributions of this work are as follows: \begin{itemize}[leftmargin=*, labelindent=0pt]
    \item We explore a new attack surface for Mixture-of-Experts based transformers in modern efficient AI, \mymethod, which leaks the user's private inputs and outputs via the execution footprints of dynamic activation of experts.
    \item We propose four side-channels against MoE transformers deployed on both CPUs and GPUs. Based on them, we propose four privacy attacks on MoE-based large language models and vision-language models.
    \item Our evaluations show that the MoE-based model execution patterns leak the keywords of user input prompts, all system response tokens, and attributes of user input images, revealing serious privacy concerns of the MoE structure.
    \item We implement an end-to-end attack using the proposed side-channels against the DeepSeek-V2 Lite model on an AMD Ryzen Threadripper Pro CPU (12-cores) and NVIDIA RTX A6000, demonstrating \mymethod's effectiveness in a real system environment.
\end{itemize}

\section{Background} \label{sec: moe-description}

\subsection{Mixture-of-Experts Transformers}
The concept of MoE was first introduced in early 90s~\cite{jacobs_adaptive_1991}, and has become popular in recent transformer designs for efficient inference~\cite{dai_deepseekmoe_2024}.
During the training phase of MoE transformers, multiple expert modules are trained, each of which can specialize in a certain task (e.g., code generation) or domain (e.g., healthcare).
During the inference process of these models, a small set of expert modules is selected by an expert router to be activated for each token.
Assume an MoE-based feed forward network (FFN) layer $l$ has $M$ experts, denoted as $\mathcal{E}$=$\{e_1, e_2, \dots, e_M\}$, the input hidden state of token $x_t$ to this MoE layer is $u_t$.
The expert router is usually a gating function that determines which expert module is the suitable `expert' based on an affinity function $\boldsymbol{\phi(u_t)}$=$[\phi_1(u_t), \phi_2(u_t), \dots, \phi_M(u_t)]$. 
This routing function will assign an activation expert sequence to the given token, denoted as $\boldsymbol{S}_t = [S_{1,t},S_{2,t},\dots,S_{M,t}] \in{\{0, 1\}^M}$ where 
\begin{align}
{S}_{i,t} & = \mathds{1}\big[\phi_i(u_t) \in \text{Top}_k(\boldsymbol{\phi}(u_t))\big]. 
\label{eq:gate_selection}
\end{align}
Here, the affinities for all the experts are ranked, and only the top k experts are activated for the $t$-th token. We denote the set of all activated experts for the $t$-th token as $\mathcal{I}_t =\{i|S_{i,t}=1\}$. Then the importance scores $g_{i,t}$ is weighted using softmax over $\mathcal{I}_t$:
\begin{align}
    g_{i,t} &= \begin{cases}
\displaystyle \frac{\exp(\phi_i(u_t))}{\sum_{j\in \mathcal{I}_t} \exp(\phi_j(u_t))}, & \text{if } i\in \mathcal{I}_t \\
0, & \text{otherwise}
\end{cases} \label{eq:gate_softmax}
\end{align}
Hence, the output of the MoE layer with the input token $x_t$, $\boldsymbol{h}_t$, can be expressed as a weighted sum of the top $k$ experts' computations: 
\begin{align}
{h}_t &= {u}_t+ \sum_{i=1}^{M} g_{i,t} \cdot e_i({u}_t) \label{eq:ht_update}
\end{align}
where the experts are computed with residual links to avoid gradient vanishing.
As only the top $k$ experts are activated, when $k\ll M$, the computation is significantly reduced compared with a dense model.
On the other hand, during an MoE-based model training, there are more parameters in experts to optimize, thus increasing the capacity of large models with intact inference cost, according to the scaling law.
Nowadays, more and more modern LLM services utilize MoE architectures,
among which both the DeepSeek~\cite{dai_deepseekmoe_2024} and Llama 4~\cite{meta2025llama} are open-source, offering significant implementation details for adversaries. 
Evaluating privacy vulnerabilities and potential breaches on these architectures is of paramount importance for ensuring the security and privacy of modern AI services.

\subsection{Architecture of DeepSeekMOE}
In this work, we focus on one of the most widely used MoE structures, DeepSeekMoE~\cite{deepseek-ai_deepseek-r1_2025,dai_deepseekmoe_2024}, which offers open-sourced models in multiple versions to accommodate diverse tasks and efficient deployments (e.g., DeepSeek-V2 Lite for single-GPU or CPU systems).
Unlike other MoE architectures that assign experts at a coarse granularity (e.g., per paragraph or sentence), DeepSeekMoE employs \textbf{finer-grained expert segmentation}~\cite{dai_deepseekmoe_2024}, which splits the FFN layers into a high number of specialized expert modules and trains with load balancing loss~\cite{shazeer_outrageously_2017}  for increased flexibility and precision but also an increased input-dependency.
In DeepSeekMoE, the router 
selects the top-$k$ experts, where $k$ is a customizable parameter (e.g., $k=6$ in DeepSeek-V2 Lite with $M=64$). These experts then execute sequentially in a for loop, as shown in Code Snippet~\ref{lst:moe_infer_simplified} (Appendix~\ref{app: moe-implementation}).
The potential privacy leakage arises in this loop phase, where different experts 
have different loads of tokens, incurring different computation times  (as shown on CPUs) or requiring different computing resources (as shown on GPUs). 
In this work, we establish four different side-channels against DeepSeekMoE (or similar MoE designs) to explore privacy breaches of model users.

\yf{rephrase the previous line} 
\yf{the previous two lines need to be adjusted to be more logical. is the second line redundant?}
\rd{revised this paragraph}

\section{\mymethod--A New Attack Surface on MoE} \label{sec: method}

\begin{figure*}[t]
    \centering
    \includegraphics[width=0.99\linewidth]{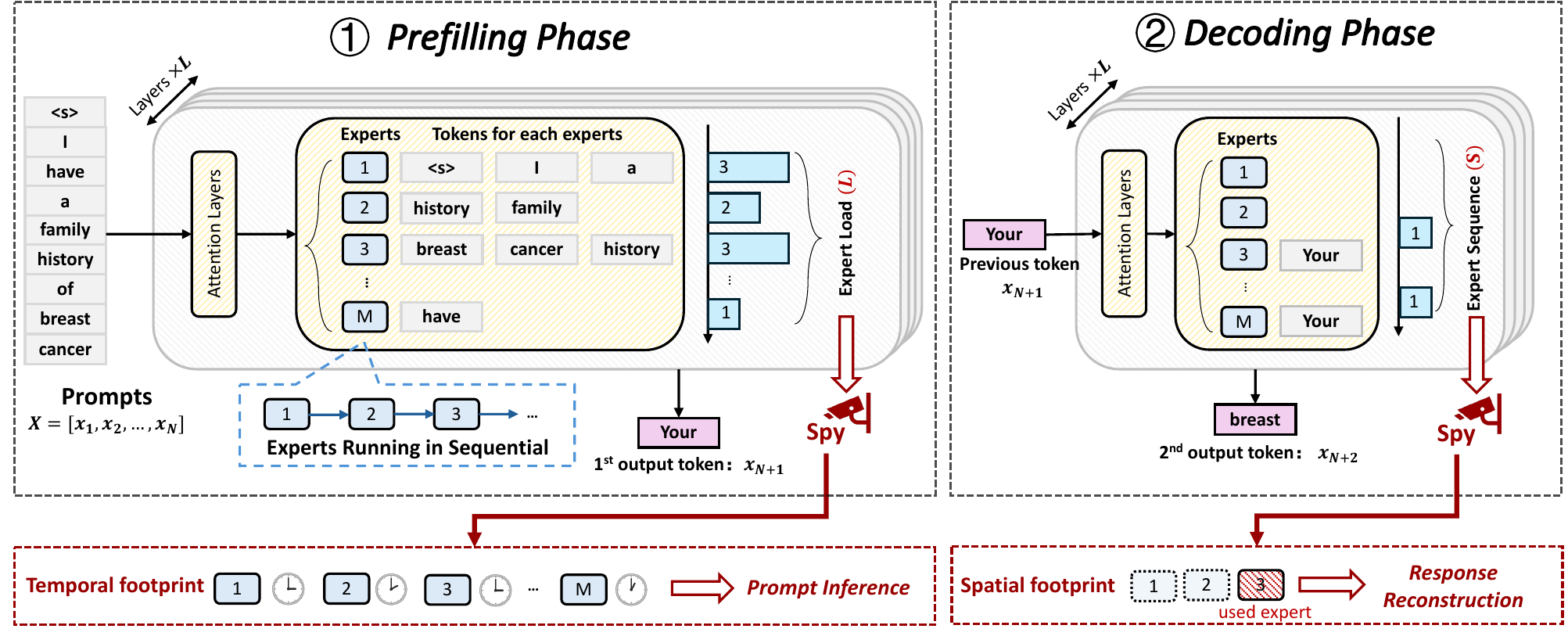}
    \vspace{2mm}
    \caption{\textit{Vulnerabilities of a Decoder-only LLM with MoE FFNs.} During the prefilling phase, the model works on all input tokens (prompts) and assigns them to different experts, where the attacker spies on the experts load to infer the user's prompts; during the decoding phase, the model processes only one input token with selected experts activated, and the spy monitors the expert sequence to reconstruct the LLM system's outputs.}
    \Description{Attack Framework}    
    \label{fig: attack framework}
\end{figure*}

Based on the intuition that dynamic expert activations in modern MoE architectures are input-dependent~\cite{yona_stealing_2024}, 
we raise three research questions as follows to understand their privacy risks.

\begin{itemize}[leftmargin=*, labelindent=0pt]
    \item \textbf{RQ1:} What components of the MoE architecture are most vulnerable, i.e., leaking sensitive information?
    \item \textbf{RQ2:} How can we effectively monitor these vulnerable components and retrieve the information leaked?
    \item \textbf{RQ3:} What privacy attacks can the leaked information enable?
\end{itemize}

\subsection{Threat Model} \label{sec: threat model}

\noindent\textbf{Victim and Adversary's Roles:}
\update{We assume that the victim is the legitimate user of a local or cloud-based MoE-LLM/VLM service. 
In the local scenario, the victim is represented by a process handling user requests, while the adversary is a malicious process (e.g., malware) operating on the same machine and sharing hardware resources with the victim. 
In the cloud setting, the victim corresponds to a virtual machine instance processing user queries, and the adversary is a malicious co-tenant on the same physical server, thereby sharing underlying resources.
We also assume that the entire system is running under nominal operating conditions with a typical workload.}

\noindent\textbf{Adversary's Goal:}
The adversary’s primary objective is to covertly extract private information from users of MoE-based models, including both their inputs and the model outputs. Specifically, in Section~\ref{sec: attacks}, we present four distinct attack scenarios targeting MoE-based LLMs~\cite{deepseek-ai_deepseek-v2_2024, jiang_mixtral_2024, meta2025llama} and VLMs~\cite{wu_deepseek-vl2_2024}. The attacks range from inferring specific input attributes (e.g., medication history) to completely reconstructing the model outputs (e.g., revised emails).

\noindent\textbf{Attacker's Capabilities:}
The attack contains two main parts: side-channel monitoring and attack model profiling.
First, we build four different side-channels across various hardware platforms, requiring different adversary capabilities. 
For CPU-based side-channels, we assume the attacker co-locates with the victim on the same CPU core and also shares the common MoE model library (e.g., the \textit{transformers} Python library). 
For GPU-based deployments, the attacker should have access to the GPU performance counters (e.g., via the Nsight system\footnote{\url{https://developer.nvidia.com/nsight-systems}}) and also share a common MoE model library with the victim.
Second, all attacks in Section~\ref{sec: attacks} rely on a profiling approach.
During the profiling phase, the adversary collects the expert execution footprints directly via code instrumentation (see Section~\ref{sec: input-dependent}) by querying the victim model with \update{public} controlled datasets.
Privacy attack models are trained used these true execution traces. 
\update{Note that in this phase, there is no side-channel information involved.}
In the subsequent attack phase, the victim's execution is monitored via various side-channels to produce side-channel traces, which are converted to execution footprints for attacks (inference).

\yf{why do you need so many places for the threat model? better consolidate}
\rd{revised.}

We answer the first question ($RQ1$) by recognizing the input-dependent expert router as a vulnerable component.

\noindent\textit{\textbf{A1: The MoE system may disclose sensitive information through input-dependent, spatio-temporal activation patterns in its expert module routing networks.}}

\subsection{Input-Dependent Expert Activation} \label{sec: input-dependent}
MoE systems reduce the computational cost by selectively activating only a small subset of experts through a gating network, while this input-dependent activation might lead to privacy breaches on users' inputs and outputs.
Let the user's prompt be composed by:
    $\boldsymbol{X}$=$[x_1, x_2, \dots, x_N]$,
where each prompt element $x_t$ is a token.
Figure~\ref{fig: attack framework} depicts the generation process of a decoder-only LLM, which is split into two phases: \textit{Prefilling Phase} and \textit{Decoding Phase}.

\noindent\textbf{Prefilling Phase.}
During this phase, the model takes the entire prompt $\boldsymbol{X}$ as input and generates the first output token. 
The computational complexity of an MoE layer is $O(Nk)$, where $k$ is the number of activated experts per token. 
As shown in Figure~\ref{fig: attack framework}, an MoE system—such as DeepSeekMoE~\cite{dai_deepseekmoe_2024}—executes the MoE FFN layers $M$ times sequentially.
As described in Section~\ref{sec: moe-description}, the MoE layer first aggregates tokens by their assigned experts and then runs each expert sequentially.
Therefore, we cannot extract the activated experts 
for individual tokens but only a more coarse-grained overview of how many of the $N$ input tokens each expert processes (see the left part of Figure~\ref{fig: attack framework}).
Hence, private information of interest in this phase is \textbf{Expert Load}, $\boldsymbol{L}$=$[L_1,...,L_M]$, 
which is defined as:
\begin{equation}
    {L_i} =  \sum_{t=1}^{N} S_{i,t}
    \label{eq: expert load}
\end{equation}
where each load $L_i$ indicates \textit{how many prompt tokens end up being processed by expert $e_i$ during the first output token generation}.
\yf{the definition of gating function is confusing, different for equations (2) and (4)} \rd{revised in current version}
\yf{shouldn't this equation for individual $L_i$??} \rd{yes}
\noindent\textbf{Decoding Phase.} 
After generating the first output token, the MoE-based model continues to generate subsequent tokens $[x_{N+1},, \dots]$ until the generation of the EOS token, but in a much more efficient way than the prefilling phase. As shown in the right part of Figure~\ref{fig: attack framework}, it leverages a Key-Value(KV) Cache~\cite{pope2023efficiently} to reduce the computational cost, where the model only needs to compute the attention and MoE feed-forward for the \emph{new} input token (i.e., the previous output token), while reusing the cached KVs for all earlier tokens. 
Consequently, at each decoding iteration, only a few experts are activated—a sparse activation pattern that correlates strongly with both the newly generated output token and the input token.
We denote this activation pattern of what experts are activated for token $t$ as the \textbf{Expert Sequence}, which shares the same definition as Equation~(\ref{eq:gate_selection}) but starting from the decoding phase ($t>N$):
\begin{equation}
    \boldsymbol{S}_t =\{S_{1,t}, S_{2,t},...,S_{M,t}\}, \text{ when } t > N.
    \label{eq: expert sequence}
\end{equation}
which is a binary sequence (with sparse 1's) and captures the exact sequence of experts used for each newly generated token.
All of our attack methods are based on these two MoE-related leakages: 
the \textbf{Expert Load} trace $\boldsymbol{L}$ from processing the user’s prompt in the prefilling phase, and the \textbf{Expert Sequence} $\boldsymbol{S}$ during decoding. 

\subsection{Sensitivity of Expert Load/Expert Sequence to User Data} 
As the expert load $\boldsymbol{L}$ and expert sequence $\boldsymbol{S}$ vary adaptively, our next step is to demonstrate that these variations are directly correlated with sensitive information contained in user inputs.
To this end, we perform a T-test~\cite{gilbert2011testing} to determine whether the MoE-related activation signals exhibit statistically significant differences between groups of input samples with distinct sensitive attributes. 
Taking DeepSeek-V2 Lite~\cite{deepseek-ai_deepseek-v2_2024} as an example, we directly record both the expert’s load $\boldsymbol{L}$ and expert sequence $\boldsymbol{S}$ during model inference by software profiling methods. 
We leverage a synthetic healthcare dataset with different types of illnesses and use these texts as input prompts and then collect $\boldsymbol{L}$ for each illness and $\boldsymbol{S}$ for each output token.
Specifically, we choose two pairs of the illness sets (each from $100$ different input prompts) and two pairs of output token sets (each contains $100$ expert sequences) and run a T-test.

\begin{figure}[t]
    \centering
    \includegraphics[width=0.99\linewidth]{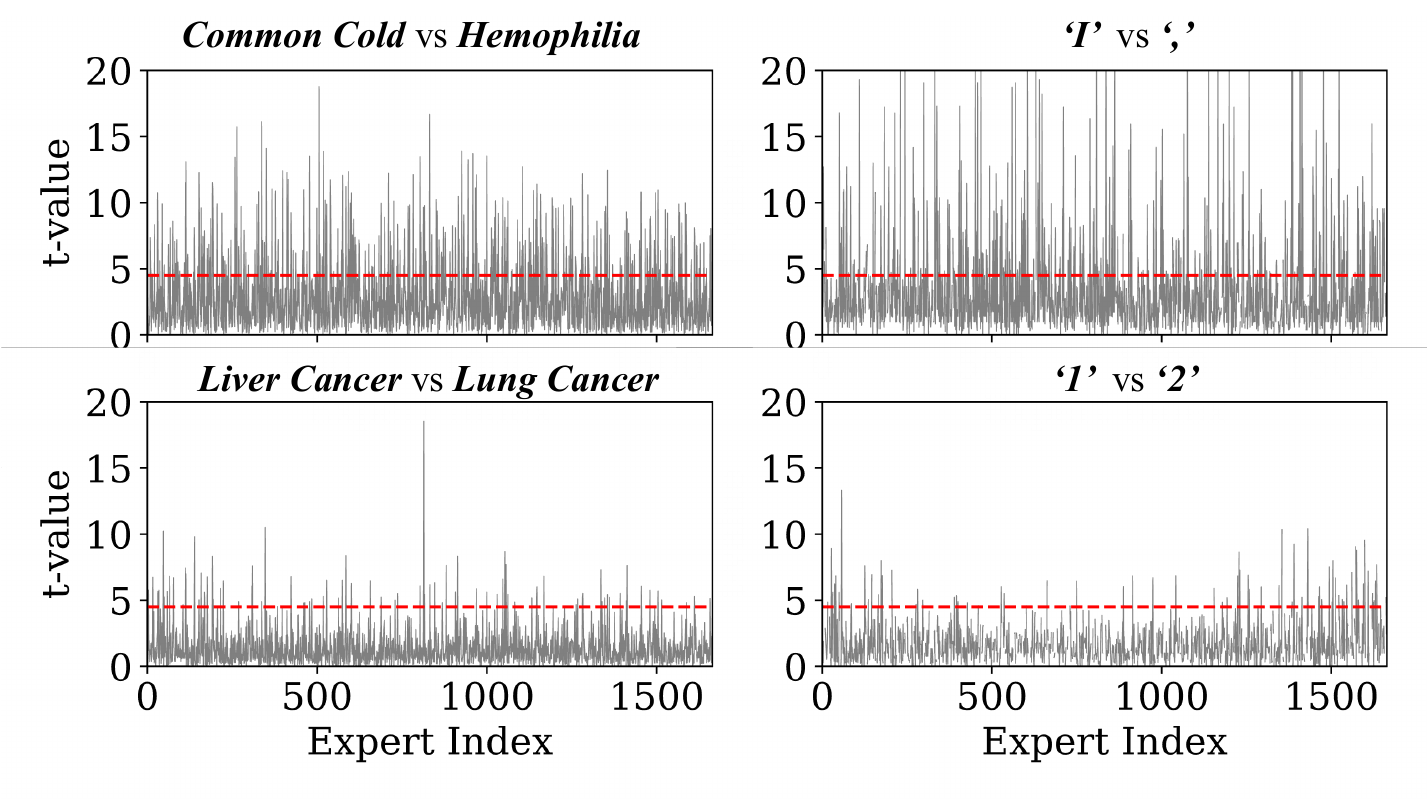}
    \caption{\textit{T-test Results of Input Prompt (left) and Output Tokens (Right)}. {\color{BrickRed} red} dash line is for t-value$=4.5$, and T-value above it indicates statistical significance. The x-axis is the expert index. (There are a total of $64\times26$ experts for the DeepSeek-V2 Lite model with 26 layers.)}
    \label{fig:t-tests}
    \Description{T-Test Results}
\end{figure}

When examining the T-test results, we primarily focus on two aspects: (1) whether there exists a statistically significant difference (i.e., t-statistics > 4.5), and (2) the portion of experts exhibiting leakage (referred to as the `leaky population'). 
In Figure~\ref{fig:t-tests}, the left two sub-figures show that distinct illness categories consistently yield statistically significant variations in the expert load $\boldsymbol{L}$. 
Specifically, our T-test is on different healthcare topic pairs (`Common Cold' vs `Hemophilia' and `Liver Cancer' vs `Colon Cancer').
For the first pair (`Common Cold' vs `Hemophilia'), there are $353$ experts that show a statistical difference;  while for the second pair of two types of cancer, only $85$ experts show a significant difference.
This phenomenon verifies that when two topics of input prompts are more semantically distinct (such as these two very different illnesses in the first pair), more experts in their expert load traces will show a significant difference.

The right two sub-figures of Figure~\ref{fig:t-tests} indicate that different output tokens also yield distinct expert sequence $\boldsymbol{S}$.
We evaluate two pairs of output tokens. 
Two distinct output tokens (`I' and `,') involve $333$ experts with  significantly different activations, while two similar ones (digits `$1$' and `$2$') involve less distinct experts ($87$).
These preliminary results underscore how observing an MoE model’s internal expert usage, including the expert load and expert sequence, can reveal information about the user’s inputs and model outputs.
Next we delve deeper into these correlations to explore the broader implications for privacy breaches, and evaluate how the `side-channel' (getting the activation patterns stealthily via architectural side-channels) closely tracks the `direct channel' (observing the execution footprints directly by instrumenting the model code).

\section{Side-channels on MoE Systems} \label{sec: SCA}
We address the second research question $RQ2$ as follows. 

\noindent\textit{\textbf{A2: Adversaries exploit architectural side-channels on model deployment platforms to extract sensitive information.}}

\begin{table}[h]
\centering
\caption{Side-channels on CPU and GPU Platforms}
\label{tab: sca attak vector structure}
\resizebox{0.99\linewidth}{!}{
\begin{tabular}{l|c|c}
\hline
\multirow{2}{*}{\textbf{Execution Pattern}}& \multicolumn{2}{c}{\textbf{Side-channels}} \\ \cline{2-3}
& \textbf{CPU} & \textbf{GPU} \\ \hline
\textbf{Expert Load $\boldsymbol{L}$} & 
\makecell{Cache Occupancy (L1 + L2) \\ \textit{Section~\ref{sec: occupancy channels}}} & 
\makecell{Performance Counter \\ \textit{Section~\ref{sec: performance counter}}} \\ \hline
\textbf{Expert Sequence $\boldsymbol{S}$} & 
\makecell{Pageout + Reload \\ \textit{Section~\ref{sec: pageout+reload}}} & 
\makecell{TLB Evict + Reload \\ \textit{Section~\ref{sec: evict+reload}}} \\ \hline
\end{tabular}
}
\end{table}
In this section, we discuss how to obtain model execution footprints via various side-channels. 
As summarized in Table~\ref{tab: sca attak vector structure}, we consider two deployment platforms, CPU and GPU, and propose four different side-channels to obtain the expert load trace $\boldsymbol{L}$ and expert sequence $\boldsymbol{S}$, respectively.
Note that we take the implementation of the DeepSeek-V2 Lite model as an example for side-channel analysis, while other MoE models (Section~\ref{sec: experiment setup}), not necessarily open-source, are also susceptible to similar side-channel analysis. 
These side-channel traces will be analyzed to retrieve the execution footprints, which are critical for privacy attacks in Section~\ref{sec: attacks}.

\subsection{SCA for MoE on CPU}\label{4.1}

\begin{figure}[h]
    \centering
    \includegraphics[width=0.99\linewidth]{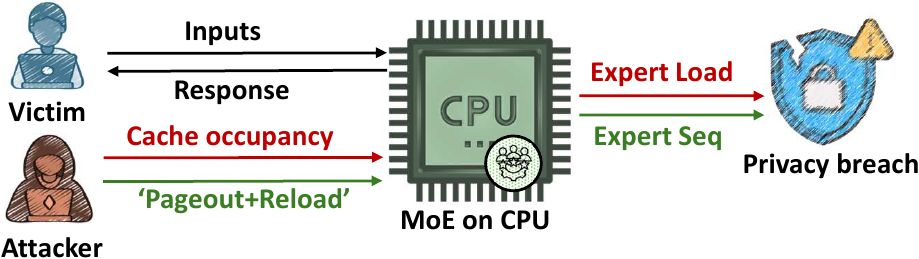}
    \vspace{1mm}
    \caption{\textit{Overview of side-channels of MoE systems on CPUs.} The attacker retrieves the expert load $\boldsymbol{L}$ with cache occupancy channels and the expert sequence $\boldsymbol{S}$ with pageout+reload.}
    \Description{Cache Attack Flow}
    \label{fig: CPU sca}
\end{figure}


To retrieve the MoE footprints on a CPU, we propose two side-channels, as shown in Figure~\ref{fig: CPU sca}: \textbf{Cache Occupancy Channels} leverage the Level-1 instruction cache and Level-2 data cache, retrieving the expert load $\boldsymbol{L}$ via different execution time of each expert;
\textbf{Pageout+Reload} assumes a shared neural network library (i.e., Python transformers), and retrieves the sequence of activated experts $\boldsymbol{S}$, via the difference of reloading time after paging out an expert memory page followed by the victim model execution.

\subsubsection{Cache Occupancy Channels} \label{sec: occupancy channels}

\begin{figure}[t]
    \centering
    \includegraphics[width=0.99\linewidth]{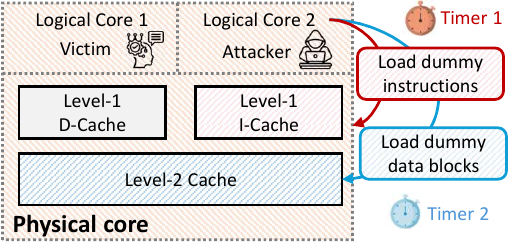}
    \vspace{1mm}
    \caption{\textit{Structure of Cache Occupancy Channels.} The co-located attacker continuously monitors the loading times for L1-Icache and L2-Dcache to infer the expert load jointly. }
    \Description{Cache occupancy channel}
    \label{fig:cache-timing}
\end{figure}

When a MoE model is running on a single CPU core, there are multiple expert modules with different token loads, executing sequentially.
As a result, the execution time of each expert is directly proportional to its token load $\boldsymbol{L}$. 
To accurately capture the execution intervals, we leverage two cache occupancy channels to identify the start and end times of each expert.
The basic idea of cache occupancy channels is to measure the victim's utilization of caches by creating contentions on the shared caches between the attacker and victim, i.e., the attacker continuously loads dummy instructions or data into a specific cache hierarchy and measures the loading time while the victim is running.  
The higher loading time indicates the higher victim cache usage.  
Figure~\ref{fig:cache-timing} depicts the two cache occupancy channels, one \textit{L1 I-Cache Occupancy Channel} and the other \textit{L2 Occupancy Channel}.

\begin{figure}
    \centering
    \includegraphics[width=0.99\linewidth]{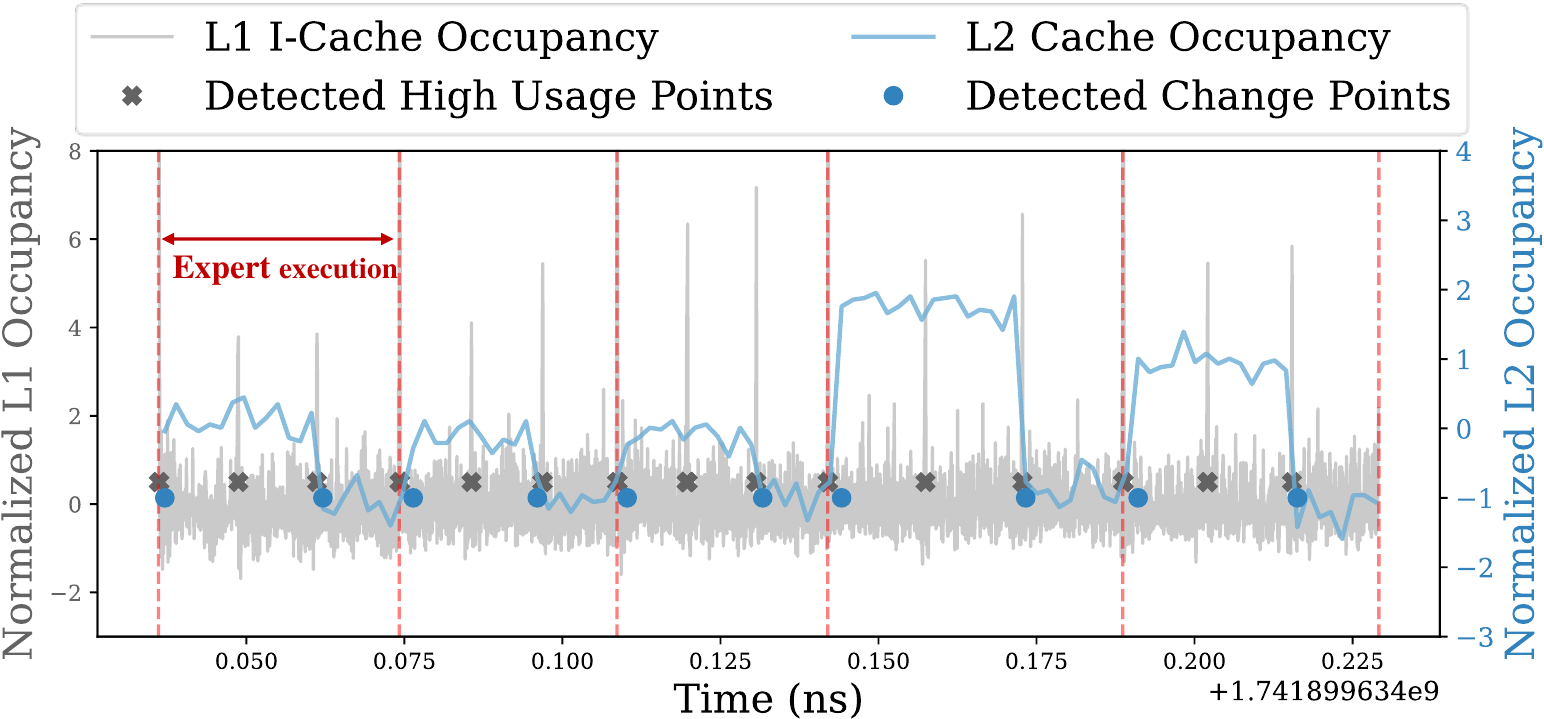}
    \caption{\textit{Raw Traces of Cache Occupancy Channels.} The {\color{gray} gray curve} shows a trace from an L1 I-Cache occupancy channel, with narrow high points marking the start of each  \textit{nn.linear} layer in PyTorch. The {\color{MidnightBlue} blue curve} shows a trace from an L2 cache occupancy channel, and the blue points indicate the change points between the first two linear layers and the third linear layer.
    The {\color{BrickRed} red dash lines} are the ground truth of start or end points of experts.}
    \label{fig: cache-occupancy}
    \Description{cache occupancy channel }
    
\end{figure}

\textbf{L1 I-Cache Occupancy Channel.}
Each expert is a computation module and consists of multiple linear layers. Each linear layer execution can be divided into three phases: \textit{module preparation}, \textit{matrix multiplication}, and \textit{end of module}.
During the module preparation and the ending phase, the layer performs a diverse set of operations--such as pointer arithmetic, memory allocations, etc, which collectively invoke a broader range of instructions and function calls in a short time, thus increasing instruction cache utilization and leading to a more pronounced ``peak'' in I-Cache usage.
However, during the intermediate phase of matrix multiplication, the computation typically involves a tightly optimized loop containing a small set of repeated instructions (e.g., fused multiply-add, vector load) for a while, resulting in lower I-Cache occupancy.
Our I-cache occupancy channel can clearly mark the linear layers' starting and ending times with a higher loading time measurement, as shown by the {\color{gray} gray curve} in  Figure~\ref{fig: cache-occupancy} (with many narrow peaks for linear layer segmentation), collected from an  AMD Ryzen Threadripper Pro CPU with a DeepSeek-V2 -Lite model running. 
As these expert modules contain three linear layers, we can see three high points for each expert on the I-cache trace.
We can calculate the distance between every three peaks, which will be proportional to the expert load $\boldsymbol{L}$.
However, the accuracy of our I-Cache occupancy channel is influenced by system noise, occasionally causing missed detections of linear layer start points—approximately 2\% in our measurements. This limitation arises from an attacker's difficulty in precisely controlling the OS scheduler's allocation of CPU time between attacker and victim processes across two logical cores. Attacker processes may intermittently be preempted by other system-level processes or interrupts, temporarily preventing timely measurement of I-Cache occupancy and thus missing some starting points. 

\textbf{L2 Occupancy Channel.}
We also leverage an L2 cache occupancy channel to assist inferring the expert load simultaneously to overcome the system noise on an L1 cache trace, as shown in the blue curve in Figure~\ref{fig: cache-occupancy}.
The L2 channel focuses on the different data cache utilization of matrix multiplication with different sizes.
For instance, in DeepSeekMoE, each expert module consists of three linear layers: an up-scaling layer, a gating layer, and a down-projection layer.
The first two layers uses a weight matrix at the same size $\mathbb{R}^{h_i\times h_j}$.
The third layer employs a weight matrix of size $\mathbb{R}^{h_j\times h_i}$ ($h_j<h_i$). 
Such linear layer computations are supported by matrix multiplication, which is optimized by a \textbf{GEMM} backend where each multiplication is divided into the multiplication of small data blocks. 
The block size is typically determined based on the weight matrix dimensions to optimize cache locality. In this context, the down-projection layer, with its larger output dimension $h_i$, tends to use larger block sizes, which improves spatial locality and thus results in lower L2 cache occupancy compared to the first two layers. 
In Figure~\ref{fig: cache-occupancy}, the {\color{MidnightBlue} blue curve} is an L2-D cache trace, and we can clearly see a higher cache occupancy for the first two linear layers and a lower occupancy for the last one.
We leverage the PELT change-point detection algorithm~\cite{killick_optimal_2012} to determine the change-points ({\color{MidnightBlue} blue \tikzcircle[MidnightBlue, fill=MidnightBlue]{0.15em}}) of these two types of data cache occupancy patterns, which mark the end of the first two layers and the endpoint of an expert.
We further use every two change points to determine the execution time of each expert.
\yf{For tianhong: what is the noise. need more elaboration.}
\tx{I added a paragragh in the end of L1 I-Cache Occupancy Channel part.}
\subsubsection{Pageout + Reload Side-channel}\label{sec: pageout+reload}

\begin{figure}[t]
    \centering
    \includegraphics[width=0.99\linewidth]{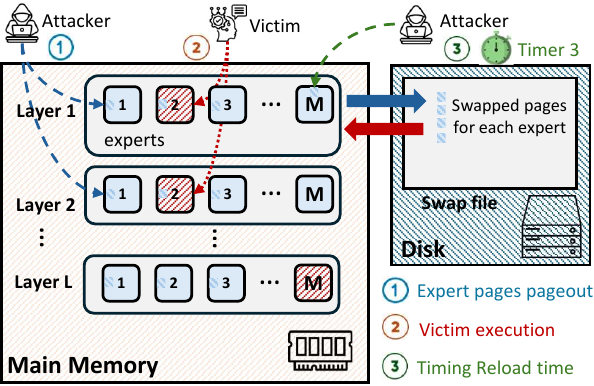}
    \vspace{2mm}
    \caption{\textit{Procedure of a Pageout+Reload Side-channel}. The attacker evicts every expert' memory page and measures the reload time to infer what experts are activated by the victim, generating an expert sequence $\boldsymbol{S}$.}
    \Description{Pageout Channel}
    \label{fig: pageout} 
\end{figure}

\begin{figure}[t]
    \centering
    \includegraphics[width=0.99\linewidth]{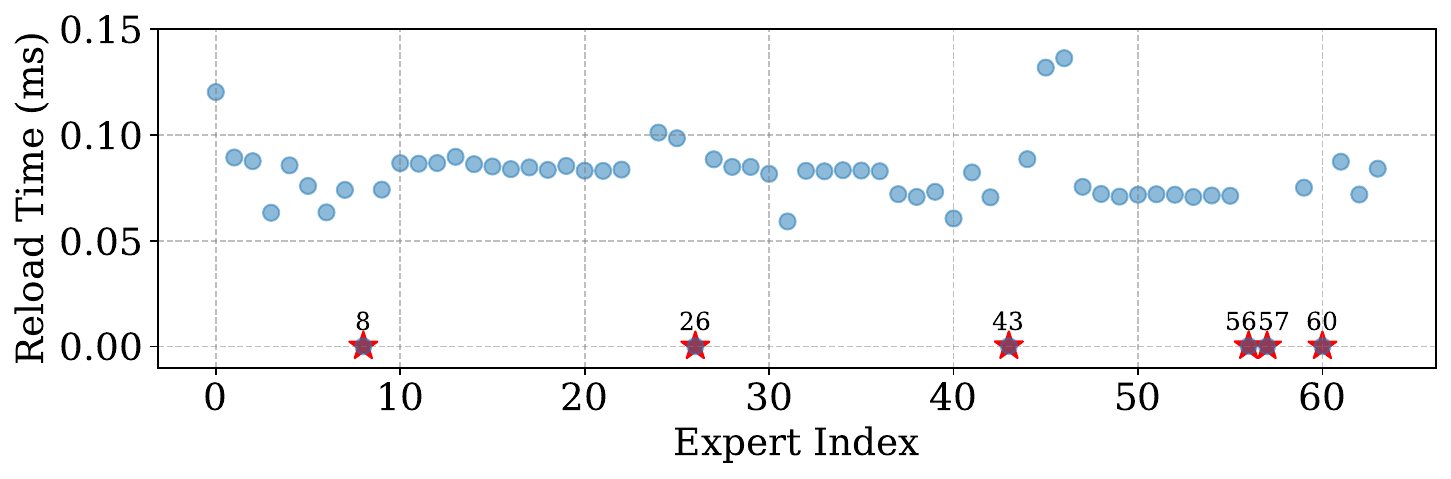}
    \vspace{2mm}
    \caption{\textit{An Example Trace from the Pageout+Reload Side-channel}. In this trace, the expert sequence $\boldsymbol{S}$ is comprised of experts with the lower reload time ({\color{BrickRed} red $\star$} in the figure). Other unused experts have a longer reload time ({\color{MidnightBlue} blue \tikzcircle[MidnightBlue, fill=MidnightBlue]{0.15em})}.}
    \label{fig: pageout-reload-trace}
    \Description{Traces of Pageout and Reload attack}
\end{figure}

To get the expert sequence $\boldsymbol{S}$ during the decoding phase of MoE systems on CPUs, we propose a \textbf{Pageout + Reload} side-channel.
We assume that the victim and the attacker are sharing the same DNN library for expert modules (i.e., shared PyTorch).  The side-channel is similar to the popular cache flush+reload side-channel~\cite{yarom2014flush+} in terms of steps, while at different architectural levels, memory pages versus cache lines. The traditional cache flush+reload side-channel would not work for MoE executions, because the cache size is limited (smaller than an expert module) and therefore for every MoE model decoding phase, the final cache state can only reflect the last activated expert. We seek alternative side-channels that can capture \textbf{all} expert activations during one decoding phase, e.g., the much larger-scale memory. 
We leverage a memory management mechanism supported by multiple OS's - pageout - that swaps out certain physical memory pages specified by the user, without changing the virtual memory mapping. 
The pageout+reload attack procedure follows three steps:  
1. Setup: the attacker pageouts one page (the first data memory page) of each expert module from the memory; 2. Victim runs (one decoding phase for outputting one token, which brings the activated experts (their pages) back into memory); 3. Reload: the attacker reloads every expert page and times it.  A short time for an expert indicates that the victim has activated this expert; while a longer time means the expert is not used for the token. 
The procedure is illustrated in Figure~\ref{fig: pageout}, where $M$ is the total number of experts. 

To pageout an expert page, we leverage a Linux system call, \texttt{madvise(addr,size,MADV\_PAGEOUT)}, which will swap out a given range of pages specified by the \texttt{addr} and \texttt{size}. 
After the setup step, all the experts have been evicted from memory. When reloading, only the experts used by the victim in the middle step will have a shorter time as they are already swapped in during MoE module execution. 
Note the attacker has to be synchronized with the victim so as to follow the 3-step procedure.  
To reduce the overhead of paging out targeted memory pages and reloading them, we only flush the first memory page of each expert.
Figure~\ref{fig: pageout-reload-trace}  presents an example Pageout+Reload measurement trace for the execution of one MoE layer of the DeepSeek-V2 Lite model.
There are six low points, indicating the six activated experts by this model, verified by the DeepSeekMoE configuration, where $k=6$ and $M=64$.

\subsection{Side-Channels for MoE on GPU}
\begin{figure}[H]
    \centering
    \includegraphics[width=0.99\linewidth]{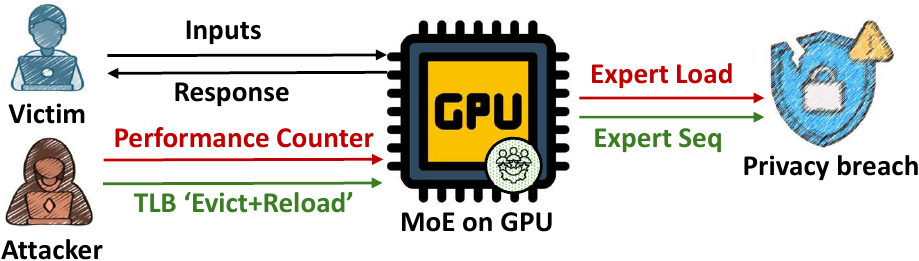}
    \vspace{1mm}
    \caption{\textit{Overview of SCA on GPU-based MoE System.} The attacker retrieves the expert load $\boldsymbol{L}$ with GPU performance counter and the expert sequence $\boldsymbol{S}$ with a TLB Evict+Reload.}
    \label{fig: gpu-flow}
    \Description{GPU flow}
\end{figure}

When the victim model is deployed on a GPU, the GPU microarchitecture, resource (memory page) management, and runtime environment are all different from CPU, and therefore the prior two side-channels for CPU do not apply any more. We propose two architectural side-channels specific for GPUs. 
We take an NVIDIA A6000 GPU as example, 
and leverage the GPU \textbf{performance counter} (via an Nvidia analysis tool -- \textit{Nsight}) to get the expert load, $\boldsymbol{L}$, and utilize a new 
\textbf{TLB Evict+Reload Side-channel} to get the expert sequence $\boldsymbol{S}$, as depicted in Figure~\ref{fig: gpu-flow}. Note accessing \textit{Nsight} does not require root privilege. 


\begin{figure}[h]
    \centering
    \includegraphics[width=0.99\linewidth]{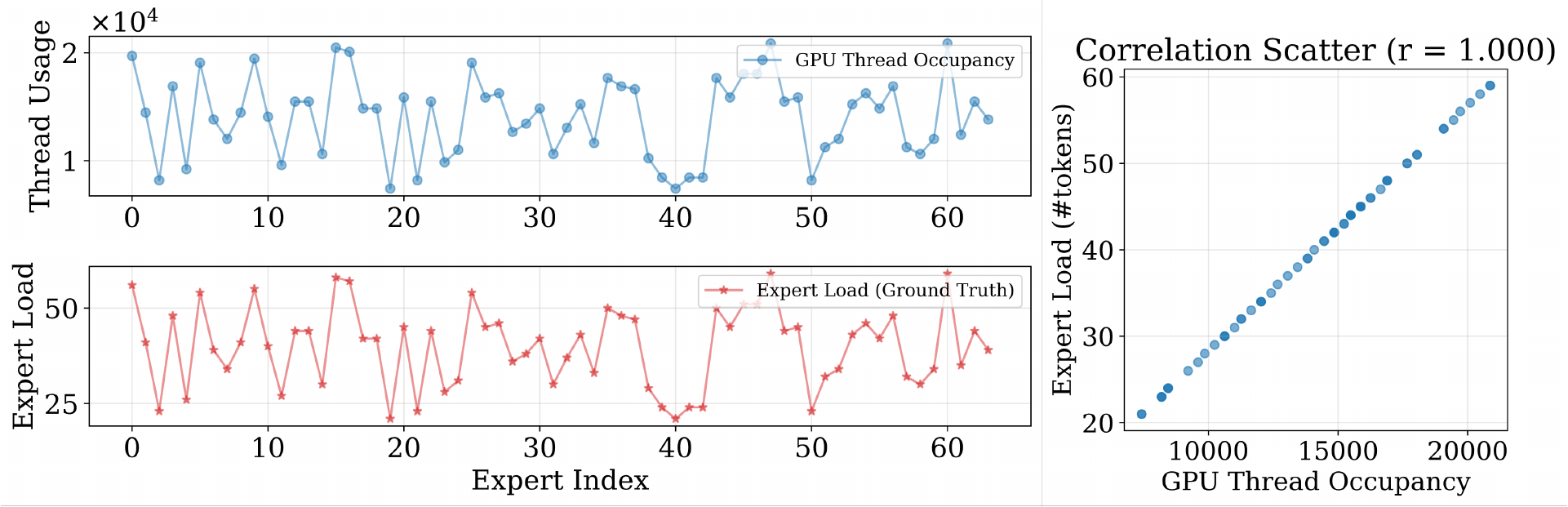}
    \caption{\textit{GPU Thread Counts and Expert Loads}. Top plot: {\color{MidnightBlue} blue curve} records the number of threads used by the activation kernels in each expert; Bottom plot: {\color{BrickRed} red curve} is the real number of tokens of each expert; Right: correlation between the thread counts and the expert token loads. }
    \label{fig:thread-occupancy}
    \Description{Thread Occupancy Attack}
\end{figure}

\subsubsection{GPU Performance Counter} \label{sec: performance counter}
Unlike sequential CPU execution, GPU features the SIMT (single instruction multiple threads) programming model, and the tokens for one single MoE expert are processed in parallel on GPUs. The number of threads involved is proportional to the number of tokens. 
Therefore, to retrieve the expert load $\boldsymbol{L}$, we use Nsight to track the number of threads each expert employs, as depicted in Figure~\ref{fig:thread-occupancy} (the {\color{MidnightBlue} blue curve}).
Specifically, we monitor the thread usage of the {\color{gray} \texttt{nn.selu}} activation function, which is called exactly once per expert. 
The number of threads observed for this kernel shows a perfect correlation (Pearson Correlation$=$1) with the expert loads ({\color{BrickRed} red curve} in Figure~\ref{fig:thread-occupancy}).

\subsubsection{TLB Evict + Reload Side-channel} \label{sec: evict+reload}

\begin{figure}[t]
    \centering
    \includegraphics[width=0.99\linewidth]{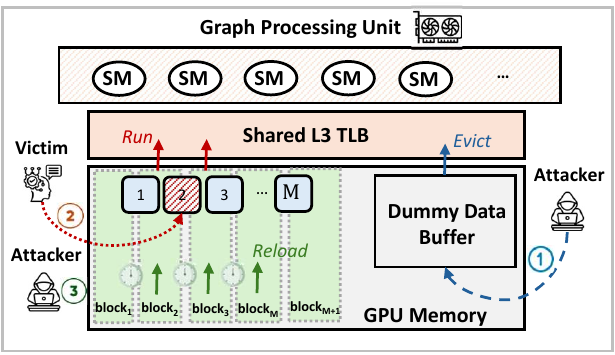}
    \vspace{1mm}
    \caption{\textit{TLB Evict+Reload Side-channel.} The attacker first loads a dummy data buffer to evict shared L3 TLB. After the victim's execution, the attacker tests the reload time of memory pages corresponding to each expert.}
    \Description{GPU TLB attack}
    \label{fig:tlb-attack}
\end{figure}

\begin{figure}[t]
    \centering
    \includegraphics[width=0.99\linewidth]{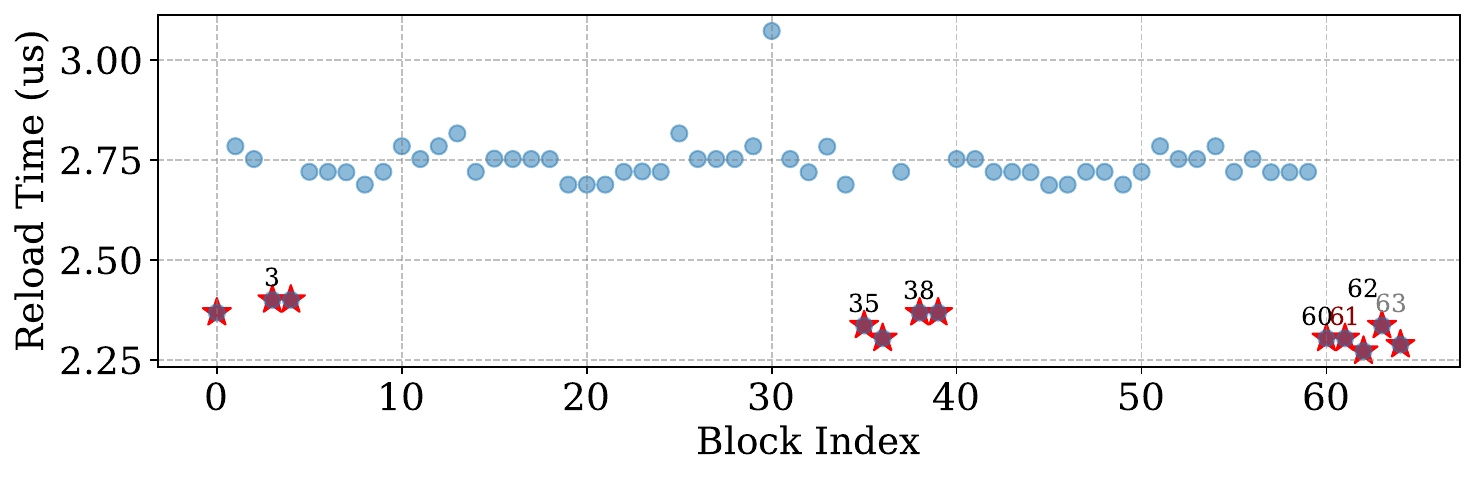}
    \caption{\textit{An Example Trace from a TLB Evict+Reload Side-channel.} The {\color{BrickRed} red $\star$} indicates the TLB entry blocks with short access time (inferred expert annotated). The {\color{MidnightBlue} blue \tikzcircle[MidnightBlue, fill=MidnightBlue]{0.15em}} denotes blocks are not in use. Ground Truth: \textbf{[3, 35, 38, 60, 62, 63]}.}
    \label{fig:uTLB-flush}
    \Description{uTLB Flush Reload Attack Traces }
\end{figure}

For GPU, there is no OS-controlled pageout mechanism. Therefore, the pageout+reload side-channel is not applicable in the GPU context. Furthermore, the page table remains unchanged once the model is deployed on a GPU. The only viable architectural component that can be manipulated as a side-channel is the Translation Lookaside Buffer (TLB), a cached portion of the page table.  
Previous works have reverse-engineered Nvidia GPUs' TLB, and established covert channels~\cite{nayak2021mis, di2021tlb, zhang2023t}.
According to~\cite{zhang2023t}, Nvidia GPUs implement a three-level TLB hierarchy: L1 TLB is private to each Streaming Multiprocessor (SM), L2 TLB is shared among a few SMs, and L3 TLB is shared across all SMs. One entry in the L3 TLB maps 16 contiguous virtual memory pages, and when huge-page of 2MB is enabled—such as the deployment of DeepSeek-V2 on Nvidia A6000—a single TLB entry can cover 32MB of contiguous virtual memory. We define the 32MB memory region as a \textbf{TLB entry block}.
We choose the shared L3 TLB to build a GPU \textbf{TLB Evict + Reload} side-channel to retrieve the expert sequence $\boldsymbol{S}$ of MoE system execution. Figure~\ref{fig:tlb-attack} shows the procedure. \tx{I checked the papers, there're actually no TLB-based side-channels on gpu, only covert channels} 

First, we evict the shared L3 TLB by accessing a large dummy buffer. Based on reverse-engineered results~\cite{zhang2023t}, an L3 TLB typically covers up to 2GB of virtual memory. 
Therefore, we allocate and access a 4GB buffer to ensure full eviction of TLB entries before the victim's execution.
After that, the victim model generates one output token involving expert routing and execution. 
The attacker then reloads a representative weight parameter from each \textbf{TLB entry block} and measures the access time. A shorter access time (TLB hit) indicates that the expert was accessed recently (i.e., `1' in the expert sequence), while a longer one (TLB miss) suggests the expert was not used by the victim.

To understand the access patterns, we analyze how experts are laid out across TLB entry blocks. In DeepSeek-V2 Lite, each MoE layer contains 64 experts. These experts are distributed across 65 TLB entry blocks, with each expert occupying two blocks ({\color{ForestGreen} green blocks} in Figure~\ref{fig:tlb-attack}). Specifically, for expert $i$, its second half shares the same memory block with the first half of expert $i+1$. This results in overlapping placement, where each intermediate TLB entry block (blocks 1 to 63) is shared by two adjacent experts. The first and last TLB entry blocks (blocks 0 and 64), however, are shared with non-MoE components (e.g., gating layers), which are always accessed during decoding. 
During the reload phase, we select one parameter from each of these blocks (i.e., the first parameter of expert 0 through expert 63, and the last parameter of expert 63) and measure the memory access time (Figure~\ref{fig:uTLB-flush}).
The lower points indicate TLB hits, indicating the block was used by victim ({\color{BrickRed} red $\star$}). 

To infer expert sequence, we declare an expert as active only if both of its corresponding TLB entry blocks (i.e., shared and individual) show TLB hits. While this method works in general, it makes errors when multiple adjacent experts are activated, due to block sharing. For instance, in Figure~\ref{fig:uTLB-flush}, the ground truth of active experts is $[3, 35, 38, 60, 62, 63]$. However, because blocks 61 through 65 all show TLB hits, we mistakenly include expert 61 in the recovered sequence, resulting in an inferred set of $[3, 35, 38, 60, 61, 62, 63]$.
This block-sharing ambiguity introduces a 3.4\% error rate in expert inference (see Section~\ref{sec: sca-performance}). Nevertheless, our attack successfully reveals the majority of the expert sequence $\boldsymbol{S}$ for attacks, demonstrating the effectiveness of L3 TLB side-channels.
\section{Proposed Attacks} \label{sec: attacks}
To answer the \textit{RQ3}, we propose multiple privacy attacks as follows.
\textit{\textbf{A3: By analyzing these side-channel leakage, attackers infer users’ inputs and stealthily reconstruct the system’s responses.}}

Based on the execution footprints, an adversary can launch privacy attacks. We propose four different attacks, including \textbf{Prompt Inference Attack} (PIA) and \textbf{Response Reconstruction Attack} (RRA) for LLM, and \textbf{Visual Inference Attack} (VIA) and \textbf{Visual Reconstruction Attack} (VRA) for VLM.
In Appendix~\ref{app: attack diagrams}, we show diagrams for each attack.

\subsection{Prompt Inference Attack (PIA)} \label{attack: PIA}
\noindent\textbf{Scenario Description.}
We propose an attack that exploits coarse-grained expert load traces obtained from the model prefilling phase. 
Instead of recovering the complete input, the attacker targets extracting sensitive attributes (for arbitrary inputs) or key phrases (for inputs following known prompt templates). 
For instance, an attacker infers the specific illness mentioned in a health-related input. The attacker first profiles the victim's MoE model with a controlled dataset that spans a wide range of sensitive information. If the inputs follow known prompt templates, prior knowledge of templates further improves extraction of privacy-revealing keywords.

\noindent\textbf{Attack Methodology.}
The attacker trains a classification model over a large output space (e.g., 116 common illnesses in our evaluation) with the MoE expert load patterns as input, i.e., the token distribution across experts. During the attack phase, the adversary observes the expert load traces via a side-channel and predicts the illness in the query, breaching the user' privacy.

\subsection{Response Reconstruction Attack (RRA)}

\noindent\textbf{Scenario Description}
Beyond user inputs,  we introduce a new attack whereby an adversary reconstructs the entire MoE model’s output verbatim.
For example, in common LLM applications—refining personal documents like emails—an eavesdropping attacker could capture sensitive information from the model’s responses. In this scenario, the adversary leverages the expert sequences during the decoding phase to establish a one-to-one mapping with output tokens, thereby reconstructing the complete response.

\noindent\textbf{Attack Methodology}
The attacker begins by profiling the MoE model using a controlled dataset to characterize its execution (expert sequence) of decoding. During the attack phase, expert sequences across the entire decoding phase are collected by side-channels to conduct verbatim reconstruction.

\subsection{Visual Inference Attack (VIA)} \label{sec: via}

\noindent\textbf{Scenario Description.}  
MoE architectures are widely used in multi-modal settings~\cite{dai_deepseekmoe_2024}. 
We present two attacks against MoE-based vision-language models that focus on recovering sensitive information from the vision modality. 
The first attack recovers coarse-grained visual attributes—such as private facial features—and may reveal the identity of the model user.

\noindent\textbf{Attack Methodology.}  
For visual attribute inference, the attacker exploits MoE signals to extract key attributes from the input image. 
To infer the identity, we assume that the attacker possesses a database of all identities and matches the recovered MoE signal against this database to determine the most likely identity.

\subsection{Visual Reconstruction Attack (VRA)}

\noindent\textbf{Scenario Description.}  
In the Visual Reconstruction Attack (VRA), the goal is to recover fine-grained visual details from MoE execution leakage. We assume that the attacker possesses partial prior knowledge of the input image (e.g., a masked version) and uses this information to guide a generative model in reconstructing the complete image. 
A representative case arises in privacy-sensitive image processing systems, e.g., a medical imaging service that applies facial masking to protect patient privacy. 
While the system displays masked results to users, its underlying model processes the original unmasked images during computation. 
This discrepancy creates an execution leakage that VRA exploits to reconstruct the original images containing sensitive information.

\noindent\textbf{Attack Methodology.}  
The attacker trains a conditional generative model that leverages the MoE signal (i.e., the expert load trace) as a conditioning input to accurately reconstruct the original image.

\section{Evaluations} \label{sec: evaluation}
\begin{table}[t]
\centering
\caption{MoE Architectures Under Test}
\label{tab:moe_model}
\resizebox{0.99\linewidth}{!}{
\begin{tabular}{l|c|c|c|c|c|c}
\hline
\textbf{Model} & \textbf{\#Params} & \textbf{Active} & \textbf{\#Exp} & \textbf{Exp Size} &\textbf{Balance} & \textbf{Type} \\
\hline
DeepSeek-V2~\cite{deepseek-ai_deepseek-v2_2024} & 16.4B & 2.8B & 64 & 8.65M &  \cmark & LLM \\
Qwen1.5-MoE~\cite{qwen_moe} & 14.3B & 2.7B & 60 & 8.65M &  \cmark & LLM\\
TinyMixtral~\cite{jiang_mixtral_2024} & 701M & 400M & 4 & 12.6M  & \xmark & LLM \\
DeepSeek-VL2~\cite{wu_deepseek-vl2_2024} & 3.37B & 1.0B & 64 & 3.00M  &\cmark & VLM\\
\hline
\end{tabular}
}
\end{table}

\begin{table*}[t]
\centering
\caption{Comparison of Prompt Inference Attack Performance Across Three MoE-based LLM Models}
\label{tab: pia-performance}
\resizebox{0.99\textwidth}{!}{
\begin{tabular}{l|cccc|cccc|cccc}
\hline
\multirow{2}{*}{Prompt Settings} & \multicolumn{4}{c|}{\textbf{DeepSeek-V2-Lite~\cite{deepseek-ai_deepseek-v2_2024}} (64 experts)}  & \multicolumn{4}{c|}{\textbf{Qwen-1.5b-MoE~\cite{qwen_moe}} (60 experts)} & \multicolumn{4}{c}{\textbf{TinyMixtral-MoE~\cite{jiang_mixtral_2024}} (4 experts)} \\
\cline{2-13}
                         & Age (8) & Gender (2)  & Blood (8) & Illness (116)  & Age (8) & Gender (2)  & Blood (8) & Illness (116) & Age (8) & Gender (2) & Blood (8) & Illness (116) \\
\hline
Unstructured (S)         &  45.6\%  &    81.1\%   &   45.7\%  &  56.5\%     &  47.5\%   &   81.4\%   &  55.7\%     &    74.4\%     &  21.6\%   &   65.6\%     &  23.7\%     &   9.50\%       \\
Unstructured (L)      &   40.0\%  &    75.6\%    &  26.6\%    &  68.8\%     &  35.8\%  &   80.1\%    &  39.8\%    &    94.5\%    &   18.3\%  &    71.3\%    &  19.7\%     &     8.20\%    \\
\hline
Templated (S)    &98.9\% &100\% &99.8\%  &99.9\%        &   98.9\%  &   99.9\%     &   99.8\%    &    100\%     &   48.9\%  &   96.4\%     &  58.3\%     &    47.3\%     \\
Templated (L)   &85.5\% & 100\% &96.1\%  & 97.9\%        &   72.7\%   &   93.3\%      &   98.0\%      &    96.0\%      &  37.6\%   &   94.1\%     &  46.7\%     &  53.8\%       \\

Cross-Template      & 83.9\% &92.0\% &72.3\%  & 95.4\%       &  78.1\%   &   98.7\%     &  76.7\%     &    96.8\%     & 21.1\%    &  53.6\%      &    29.5\%   &    16.6\%     \\
\hline
\end{tabular}
}
\end{table*}

\begin{table*}[t]
\scriptsize
    \centering
        \caption{\textit{RRA Performance with Expert Sequence} on Synthetic Prompt~\cite{thedrcat2024llm}, Medical Q\&A~\cite{thedevastator2023medicalqa}, Financial Q\&A Dataset~\cite{saeedian2023financialqa}.}
\resizebox{0.99\textwidth}{!}{
    \begin{tabular}{c|c|c|c}
    \hline
    \textbf{Dataset -- Input Prompt}     & \textbf{System Response (Ground Truth)} &  \textbf{Reconstructed Response (Attack Results)} & \textbf{ASR}   \\
    \hline
       \parbox{0.21\linewidth}{\textbf{Synthetic}~\cite{thedrcat2024llm}--Change the text into an art curator's description of a modern art exhibit, ...   
Text:
He has been the Prime Minister of this country not once, but twice, so I acknowledge his service to the people of our nation. Tony Abbott gives his victory speech at the Four Seasons Hotel, Sydney. ......}&
\parbox{0.39\linewidth}{\vspace{1mm} 
Art Curator's Description:

In the heart of Sydney's bustling {\color{BrickRed}Four Seasons} Hotel, a monumental tableau unfolds, capturing the essence of contemporary political art. The scene, a triumphant speech by Tony Abbott, Prime Minister of this nation twice over, is rendered in bold, abstract brushstrokes that {\color{BrickRed}speak} to the transformative power of leadership. The artist, known for their innovative use of color and form, juxtapos{\color{BrickRed}es} the traditional {\color{BrickRed}trappings} of political victory with the raw, emotional energy of the crowd.
\vspace{1mm}}
& \parbox{0.39\linewidth}{\vspace{1mm}
{\color{YellowGreen}
Art Curator's Description:

In the heart of Sydney's bustling {\color{BrickRed}Sixhour} Hotel, a monumental tableau unfolds, capturing the essence of contemporary political art. The scene, a triumphant speech by Tony Abbott, Prime Minister of this nation twice
 over, is rendered in bold, abstract brushstrokes that {\color{BrickRed}speaks} to the transformative power of leadership. The artist, known for their innovative use of color and form, juxtapos{\color{BrickRed}apates} the traditional {\color{BrickRed}trings} of political victory with the raw, emotional energy of the crowd.}
\vspace{1mm} } 
&93.6\% \\
    \hline
     \parbox{0.21\linewidth}{\textbf{Medical}~\cite{thedevastator2023medicalqa} - Who is at risk for Lymphocytic Choriomeningitis (LCM)? ?} &  
     \parbox{0.39\linewidth}{\vspace{1mm}
     Lymphocytic choriomeningitis (LCM) is a viral infection that is transmitted to humans from infected rodents. Lymphocytic choriomeningitis virus (LCMV) is a member of the Arenavirus family. LCM is a zoonotic disease, meaning it can be transmitted from animals to humans. The primary reservoir for LCMV is the {\color{BrickRed}house} mouse ({\color{BrickRed}Mus musculus}). However, other rodents, such as {\color{BrickRed}deer} mice, {\color{BrickRed}cotton rats}, and {\color{BrickRed}white-footed} mice, can also be infected.
     \vspace{1mm}}
  &  \parbox{0.39\linewidth}{\vspace{1mm}
  {\color{YellowGreen} Lymphocytic choriomeningitis (LCM) is a viral infection that is transmitted to humans from infected rodents. Lymphocytic choriomeningitis virus (LCMV) is a member of the 
 Arenavirus family. LCM is a zoonotic disease, meaning it
 can be transmitted from animals to humans. The primary reservoir for LCMV is the {\color{BrickRed} home} mouse ( {\color{BrickRed}Gall arei}). However, other rodents, such as {\color{BrickRed}brain} mouse, {\color{BrickRed}coal, and Black-legged cells}, can also be infected.} 
 \vspace{1mm}}& 91.5\% \\
   \hline

   \parbox{0.21\linewidth}{\textbf{Financial}~\cite{saeedian2023financialqa}--What significant invention did NVIDIA create in 1999?}&
\parbox{0.39\linewidth}{ \vspace{1mm}
 NVIDIA continues to invest in deep learning research and development, and {\color{BrickRed}their} GPUs remain a cornerstone of the deep learning ecosystem. As the technology evolves, we can expect to see even more sophisticated applications of deep learning in various fields, powered by NVIDIA's advanced hardware.
NVIDIA Corporation is a multinational corporation that designs graphics processing units for the gaming and professional markets, as well as system on a chip units for the mobile computing and automotive market. In {\color{BrickRed}1999}, NVIDIA introduced the GeForce 256 GPU, which was the world's first GPU to offer programmable {\color{BrickRed}vertex} and {\color{BrickRed}pixel shaders}. This was a significant {\color{BrickRed}advancement} in the field of computer graphics, as it allowed for more complex and realistic graphics in video games and other applications. The GeForce 256 also introduced the concept of GPU-accelerated computing, which has since become a major trend in the tech industry.
\vspace{1mm}}
& \parbox{0.39\linewidth}{\vspace{1mm}
{\color{YellowGreen}
NVIDIA continues to invest in deep learning research and development, and {\color{BrickRed}its} GPUs remain a cornerstone of the deep learning ecosystem. As the technology evolves, we can expect to see even more sophisticated applications of deep learning in various fields, powered by NVIDIA's advanced hardware.
NVIDIA Corporation is a multinational corporation that designs graphics processing units for the gaming and professional markets,
 as well as system on a chip units for the mobile computing and automotive market. In {\color{BrickRed}1989}, NVIDIA introduced the GeForce 256 GPU, which was the world's first GPU to offer programmable {\color{BrickRed}matrix} and {\color{BrickRed}vector Tr models}. This was a significant {\color{BrickRed}development} in the field of computer graphics, as it
 allowed for more complex and realistic graphics in video games and other applications. The GeForce 256 also introduced the concept of GPU-accelerated computing, which has since become a major trend in the tech industry.
}
\vspace{1mm}}
&91.9\% 
\\

   \hline
    \end{tabular}
}
    \label{tab: RRA-performance}
\end{table*}

\subsection{Experiment Setups} \label{sec: experiment setup}
\noindent\textbf{Models.} In this work, we evaluate \mymethod on multiple pre-trained MoE-based models, listed in Table~\ref{tab:moe_model}, including three LLMs, DeepSeek-V2~\cite{deepseek-ai_deepseek-v2_2024}, Qwen1.5-MoE~\cite{yang2024qwen2}, and TinyMixtral~\cite{jiang_mixtral_2024}, and one VLM DeepSeek-VL-2~\cite{wu_deepseek-vl2_2024}. 
These models differ in terms of the overall model size and MoE configuration. 
Specifically, the DeepSeek and Qwen1.5 models are examples of fine-grained MoE architectures, where each expert is highly specialized, featuring relatively small expert sizes and a large number of experts. 
Both the DeepSeek and Qwen1.5 models employ the balanced expert loss~\cite{dai_deepseekmoe_2024, yang2024qwen2}. 
In contrast, TinyMixtral---a compact version of Mixtral~\cite{jiang_mixtral_2024}---is designed for single-GPU usage, with only four large experts, of which two are activated during inference. 
All these models can be executed on either a CPU or a single GPU for inference, making the models susceptible to side-channel exploitations proposed in  \mymethod.

\noindent\textbf{Datasets.}
Multiple datasets are used to evaluate proposed attacks:
\begin{itemize}[leftmargin=*, labelindent=0pt]
    \item \textbf{PIA and RRA on LLM}: To evaluate the privacy leakage of the MoE-based language model, we generate a synthetic text dataset derived from open-sourced health records~\cite{prasad2022healthcare}. 
    We utilize a local AI model (DeepSeek-R1) to produce unstructured inputs based on templates provided in Appendix~\ref{app: template}. 
    In addition to the healthcare dataset, we also assess the RRA attack on the Synthetic Prompt~\cite{thedrcat2024llm}, Medical Q\&A~\cite{thedevastator2023medicalqa}, and Financial Q\&A~\cite{saeedian2023financialqa}, demonstrating the attack performance under varied contexts.
    \item \textbf{VIA and VRA on VLM}: To assess the privacy vulnerability of the MoE-based vision-language model, we adopt the CelebA human face dataset~\cite{liu2015faceattributes}, given that facial identities represent one of the most sensitive user privacy. 
    Each sample is annotated with 40 attributes. For the identity classification task, we select 300 celebrities (each with more than 25 images) from CelebA, where the attacker’s objective is to correctly identify the user input among these 300 individuals.
\end{itemize}
\ad{fill reference} \tx{have filled it}

\noindent\textbf{Metrics.}
We assess the success of the $PIA$ and $VIA$ attacks by measuring the percentage of correct guesses of input characteristics or identities. The performance of the $RRA$ attack is quantified as the percentage of output tokens correctly recovered using the MoE signals. 
For visual inputs to the VL model, the reconstruction quality is evaluated using both SSIM~\cite{hore2010image} and FID~\cite{heusel2017gans}.

\noindent\textbf{Platforms.}
We conduct all our experiments for attacks and side-channel analysis on an AMD Ryzen Threadripper Pro (12-Cores) with one NVIDIA A6000 (memory size 16 GB).

\subsection{Attack Performance} \label{sec: attack performance evaluation}
\subsubsection{Evaluation of Prompt Inference Attack}
To evaluate the PIA using \mymethod, we employ synthetic healthcare datasets from which we extract several key attributes that an attacker may attempt to recover, including  \texttt{age, gender, blood type}, and \texttt{illness}. These attributes have varying dimensions: 2 for gender, 8 for blood type, and 116 for illness. The age attribute is divided into 8 decade groups (e.g., 20–29, 30–39, etc.).
We consider two different types of inputs:
\begin{enumerate}[leftmargin=*, labelindent=0pt]
    \item \textbf{Unstructured inputs}: In this scenario, the user's input consists of unstructured text, such as a randomly composed illness description submitted to a healthcare LLM, which includes the key private attributes. In our experiments, we rely on a generative model to automatically generate these unstructured inputs.
    \item \textbf{Templated inputs}: Here, the model's input is formatted using one or more fixed templates, resembling formal healthcare records. 
    The keywords (private information) from the healthcare record are used as the input to fill out the template. 
\end{enumerate}
For each input type, we experiment with different input lengths: short (S), averaging 50 tokens per template, and long (L), averaging 150 tokens per template. In addition, we consider a cross-templated setting (C) where the model is trained with one or more known templates but is then tested using a different template.
For evaluations, we use 8000 traces to profile the PIA inference model and reserve 2000 traces for testing. Note that the number of profiling traces has a significant impact on the final attack performance, which will be discussed in detail in Section~\ref{sec: ablations}.

In Table~\ref{tab: pia-performance}, we present the PIA performance of \mymethod on both unstructured and templated inputs. By leveraging the expert loads for the PIA attack, the adversary can achieve a high success rate in guessing private attributes of the user input. For example, on DeepSeek-V2-Lite, we obtain a Top-1 accuracy of $56.5\%$ for unstructured inputs and $99.9\%$ for the templated inputs, compared to a baseline random guess accuracy of only $1/116=0.86\%$, thereby demonstrating a significant privacy breach via the MoE expert load information.
Furthermore, we observe that for fine-grained\update{, highly-specialized} MoE architectures (e.g., DeepSeek and QWen), it is easier to infer specific attributes such as the detailed illness information because the highly specialized experts tend to manifest distinct patterns for different illnesses. In contrast, the coarse-grained sparse design of TinyMixtral exhibits minimal information leakage.
Compared with unstructured input prompts, templated inputs generally yield higher attack success rates. The templated inputs exhibit more consistent expert load patterns, which reduces the noise from unrelated tokens. As a result, the PIA inference model can focus more effectively on the sensitive information that varies across samples. Moreover, compared to long inputs, shorter inputs achieve higher attack success rates due to increased information density—a short template contains the same number of private tokens, making the sensitive information more prominent.
\yf{if the result is for illness attribute, the performance numbers do not match with your table.}  \rd{revised.}
\subsubsection{Evaluation of Response Reconstruction Attack} \label{sec: rra performance}
We evaluate the \mymethod direct-channel RRA performance on three different datasets: Synthetic Prompts~\cite{thedrcat2024llm}, Medical Q\&A~\cite{thedevastator2023medicalqa}, and Financial Q\&A~\cite{saeedian2023financialqa}. These datasets have output vocabularies of 39,867, 20,681, and 17,832 unique tokens, respectively.
During the profiling phase, we train a Multinomial Logistic Regression model~\cite{kwak2002multinomial} using 90\% of the expert sequence data $\boldsymbol{S}$ and test on the remaining 10\%. 
Table~\ref{tab: RRA-performance} presents three sample prompt–response pairs from the DeepSeek-V2 Lite model for each dataset, along with their reconstructed responses. In both the reconstructed responses and the real system response, words shown in \textcolor{BrickRed}{red} are mispredicted, whereas those in \textcolor{ForestGreen}{green} are correctly predicted.
We report the Attack Success Rate (ASR)—the percentage of correctly predicted tokens—across all datasets, each of which achieves a high ASR (>90\%). Notably, many mispredicted tokens still carry similar semantic meaning (e.g., “house” vs. “home,” “speak” vs. “speaks,”  “advancement” vs. “development”), indicating that expert sequences leak semantic information about system’s responses.

\subsubsection{Evaluation of Visual Inference Attack}

\begin{figure}[h]
    \centering
    \includegraphics[width=0.99\linewidth]{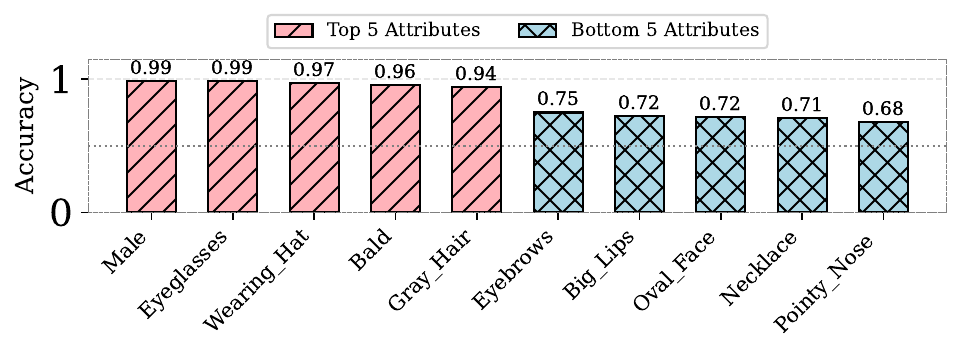}
    \caption{\textit{Performance of VIA}. We present Top-5 accurate attributes (Male, Eyeglasses, Wearing Hat, Bald, Gray Hair) in {\color{BrickRed} red bars} and Bottom-5 attributes (Arched Eyebrows, Big Lips, Oval Face, Wearing Necklace, Pointy Nose) in {\color{MidnightBlue} blue bars}. }
    \label{fig: AIA attack performance}
    \Description{Attributes inference Attack Performance}
\end{figure}

Samples in the CelebA-dataset~\cite{liu2015faceattributes} have $40$ labeled binary attributes, and the VIA attack aims to infer those attributes. The victim model is DeepSeek-VL2~\cite{wu_deepseek-vl2_2024}, with a fixed text prompt `Describe this figure' companioned with each image sample. We utilize the expert load of the model execution for the VIA.
A simple 3-layer MLP binary classifier is used with $80\%$ of MoE expert load traces as the training set and the rest as the testing set.
Figure~\ref{fig: AIA attack performance} presents the Top-5 and Bottom-5 VIA attack success rates. 
We can infer most of the coarse-grained features such as `gender', `wearing eyeglasses', etc. 
However, some detailed features, including `pointy nose', have a lower ASR.
The average VIA success rate is \textbf{$85.34\%$}.

As described in Section~\ref{sec: via}, one of the applications of VIA is for identity inference, where the attacker aims to figure out the identity of the input image among a given database of portraits.
We select $300$ identities from the CelebA dataset, with each person has more than $25$ images.
With a simple SVM classifier, the one-shot \textbf{Top-1 accuracy is $25.0\%$} and \textbf{Top-5 accuracy is $54.4\%$}, indicating that an adversary can effectively leverage MoE signals to deduce the identity of the individual depicted in the input image, thereby posing a significant risk to user privacy in the VLM model.
\ad{I do not understand the above two sentences. (1) is the accuracy for identity (which person it is) or attributes? If it is identity, why you say 'user figure's private features' instead? (2) How does the VIA helps faking authentication? Do not see the relationship here.} 
\rd{Revised. Make this statement more general.}

\subsubsection{Evaluation of Visual Reconstruction Attack}
\begin{figure}[t]
    \centering
    \includegraphics[width=0.99\linewidth]{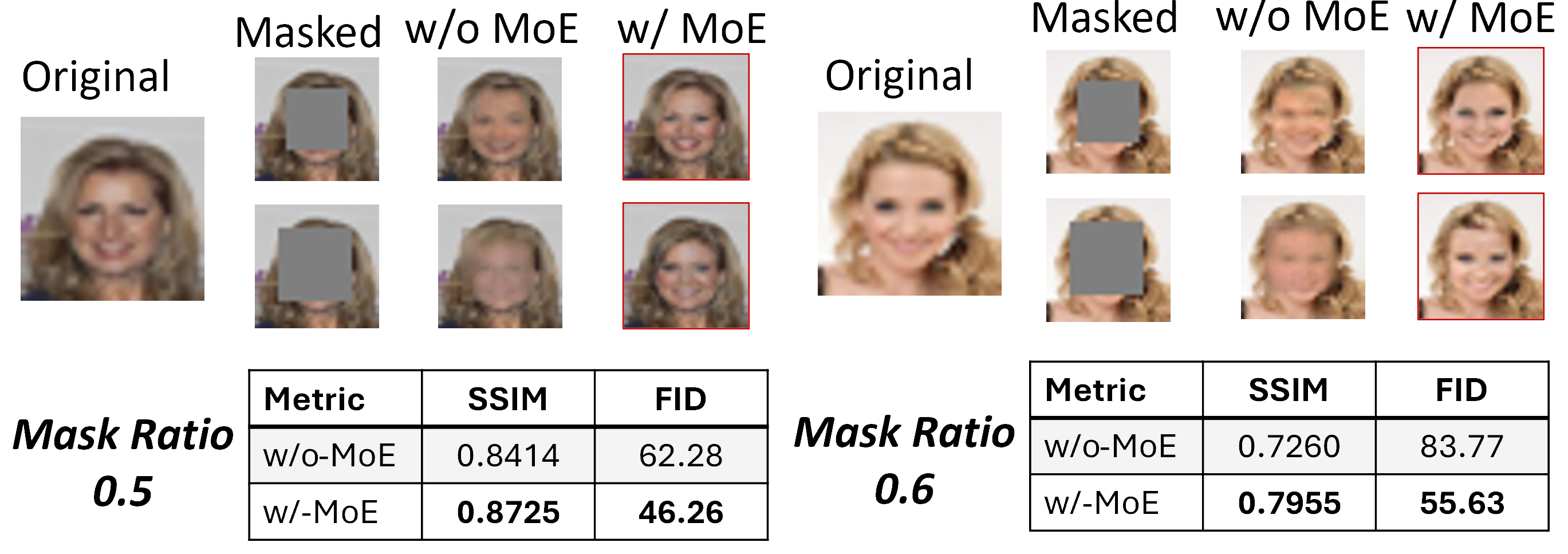}
    \vspace{1mm}
    \caption{\textit{Visual Reconstruction Performance}. Each test image is masked with two different ratios ($0.5$/$0.6$). The reconstruction performance is calculated with random condition (w/o MoE) or with Expert Load $\boldsymbol{L}$ as input condition (w/ MoE). We report two metrics for the inpainting performance, SSIM (higher the better), FID (lower the better).}
    \label{fig: vra-results}
    \Description{VRA-results}
\end{figure}

To further show the information leakage from a different modality of the \mymethod attack surface, we evaluate the VRA in an image inpainting task.
For instance, the masked image can be the output from an MoE-based DeepFake model, and the attacker's goal is to reconstruct the real user's input image.
We build a conditional GAN-based image inpainting model (more details are in Appendix~\ref{app: attack-structure}), and use the expert load $\boldsymbol{L}$ from the victim model (DeepSeek-VL2) as the generation condition. For comparison, we also reconstruct the image from a masked image with a random condition (no assistance of the MoE execution information). In Figure~\ref{fig: vra-results}, incorporating expert load information leads to improved generation performance in visual quality--the reconstruction highlighted with {\color{BrickRed} red border} preserve more facial details than the one without. 
Also, the improvement is reflected as higher SSIM scores--which evaluate the structural similarity between the original and reconstructed images, and lower FID scores, which show MoE results are more semantically similar.

\subsection{End-to-End Evaluation} \label{sec: end2end}

\subsubsection{Attack Framework}
Side-channel \mymethod attacks includes:
\begin{enumerate}[leftmargin=*, labelindent=0pt]
    \item \textbf{Profiling Phase.} The attacker collects execution footprints (the expert load and expert sequence) of the victim model on a synthetic dataset by a software method (e.g., instrumentation). 
    Then, based on the profiled execution details, the attacker fits a model for a specific attack (e.g., critical prompts, personal attributes or output tokens), as described in Section~\ref{sec: attack performance evaluation}. 
    \item \textbf{Side-Channel Trace Collection.}
    The attacker runs the side-channel attacks on the victim platform and collects traces, including the cache occupancy, reloading time, etc.
    \item \textbf{Attacking Phase--translate side-channel information to execution footprints.} 
    The attacker converts the side-channel traces to execution footprints (expert load and expert sequence) utilizing algorithms (e.g., changepoint detection), and leverages the pre-trained attack models to infer the user's secret.
\end{enumerate}

\subsubsection{Side-Channel Attack Performance} \label{sec: sca-performance}

\begin{table}[t]
    \centering
    \caption{\textit{Side-Channel Attack Translation Performance}}
    \resizebox{0.91\linewidth}{!}{
    \begin{tabular}{l|c|c}
   \hline
   \multirow{2}{*}{Performance} & \multicolumn{2}{c}{Platform}\\        
    \cline{2-3}
     & CPU    & GPU\\                    
    \hline
    \multirow{2}{*}{Expert Load  (Corr)}   &  Cache Occupancy         &  Performance Counter \\  
    \cline{2-3}
                                        &  0.901   & 0.993 \\
    \hline
    \multirow{2}{*}{Expert Sequence  (\%)} & Pageout+Reload     &  TLB Evict+Reload \\
    \cline{2-3}
                                         & 99.2  & 94.3\\
    \hline
    \end{tabular}
    }
    \label{tab: sca performance}
\end{table}

To evaluate our proposed architectural side-channels, we measure the translation accuracy from the SCA traces to the execution footprints. Specifically: 1. For the expert load $\boldsymbol{L}$, we report the Pearson correlation between the ground truth and the side-channel measurements; 2. For the expert sequence $\boldsymbol{S}$, we present the percentage of correctly identified activated experts.
Table~\ref{tab: sca performance} shows that using the combined cache occupancy channels (L1 and L2) achieves a load trace correlation of 0.9 with the ground truth. In contrast, utilizing the GPU performance counter— which directly measures the computational resources consumed by each expert module—yields a more accurate result with a correlation coefficient of 0.993.
For recovering the activated expert sequence $\boldsymbol{S}$, both the CPU-based Pageout+Reload and the GPU-based TLB Evict+Reload techniques achieve high accuracy levels of 99.2\% and 94.3\%, respectively.
The following sections will further discuss the performance of two end-to-end attack scenarios.

\subsubsection{Privacy Attacks Performance based on SCA}
In Table~\ref{tab: sca-end2end}, we report the end-to-end attack success rate for all four types of side-channel attacks. We use ``Footprints" to annotate the attack that uses the expert load and expert sequence information directly. We evaluate the overall performance using two datasets: PIA (Templated (S)) and RRA (Synthetic Prompts). For PIA attacks, both the CPU-based cache occupancy channel and the GPU-based performance counter channel exhibit robust performance, which can be attributed to the high correlation between the expert load and the corresponding cache occupancy and performance counter values. Although the occupancy channel occasionally misses some data points, the non-ideal correlation does not impact the overall attack performance much.
In contrast, the RRA end-to-end evaluation reveals that the CPU Pageout+Reload attack maintains a high accuracy, incurring only a 0.8\% drop in the attack success rate compared with the direct-channel attack. Conversely, the TLB Evict+Reload attack—which, as discussed in Section~\ref{sec: evict+reload}, suffers from mispredictions at the first and last data points, leading to notable degradation in its attack success rate. However, we still achieve an 82.5\% accuracy in output tokens, effectively revealing private information in it.



\begin{table}[h]
    \centering
    \caption{\textit{End-to-end \mymethod Attack} on DeepSeek-V2 Lite}
    \resizebox{0.7\linewidth}{!}{
    \begin{tabular}{l|c|c|c}
    \hline
    Dataset & Footprints  & CPU SCA  & GPU SCA \\
    \hline
    PIA   & 99.9\%  & 99.8\%  & 99.8\%  \\
    RRA    & 93.6\%  &  92.8\% & 82.5\%   \\
    \hline
    \end{tabular}
    }

    \label{tab: sca-end2end}
\end{table}

\subsection{Ablation Studies} \label{sec: ablations}

The profiling procedure significantly influences the performance of \mymethod-based attacks. In this section, we present ablation studies to examine how both the quantity and the quality of the profiling dataset affect attack performance.

\subsubsection{Impact of Profiling Data Scale} \label{sec: ablation-1}
\begin{figure}[h]
    \centering
    \includegraphics[width=0.95\linewidth]{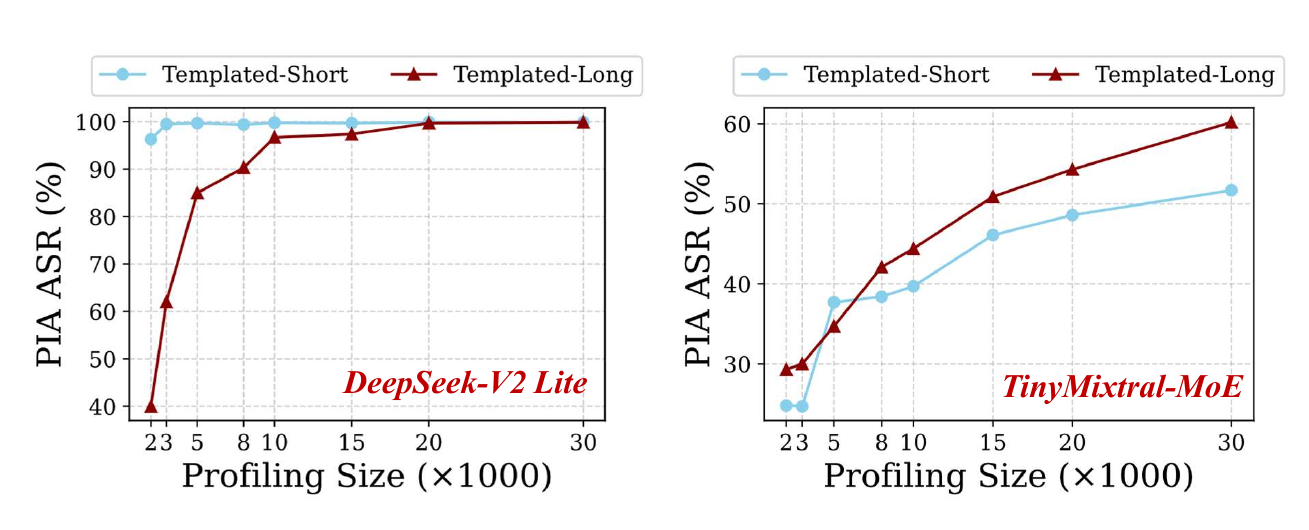}
    \caption{\textit{PIA Performance at Multiple Profiling Data Scale.}}
    \Description{Ablation-1}
    \label{fig: ablation 1}
\end{figure}

We first investigate how the number of profiling traces—i.e., the collected expert execution signals used to train the model—affects the performance of the proposed \mymethod attack. In Figure~\ref{fig: ablation 1}, we analyze the templated PIA performance for varying sizes of the profiling dataset on two architectures: DeepSeek-V2-Lite (26×64 experts) and TinyMixtral-MoE (12×4 experts). We report results using both long and short templates.
Specifically, both datasets show improved performance as the training dataset size increases. The improvement is especially pronounced for TinyMixtral-MoE, where each sample provides less expert leakage information, necessitating a larger dataset to achieve convergence. However, since collecting profiling traces incurs implementation cost, there is a trade-off between attack performance and cost, which we discuss in Section~\ref{sec: attack cost}.

\subsubsection{Impact of Profiling Context} \label{sec: ablation-2}
Except for the scale of profiling data, the context of profiling data also has a strong effect on the profiling performance. 
In Table~\ref{tab: pia-performance}, we assess the cross-template attack performance of the PIA attack. 
When we use different templates during evaluation, the attack performance will drop. 
Further, if we have random context (i.e., the unstructured inputs), the performance drops significantly.
Other than PIA, we also analyze the cross-domain performance of RRA, i.e., when we use the news dataset for profiling can we still reconstruct the system responses of a healthcare request?
As shown in Table~\ref{tab: rra-cross}, we report the ASR of RRA across the three datasets given in Section~\ref{sec: rra performance}.
When profiling with different contexts, the ASR of RRA drops due to missing the knowledge of domain-specific tokens. 
However, one can also leverage multiple general datasets to improve the attack performance (i.e., using all datasets in Table~\ref{tab: rra-cross}). 
\ad{Need to specific as to row versus column, which is profiling source, and which is attack target (columns are profiling sources?)}\rd{revised}

\begin{table}[t]
    \centering
    \caption{\textit{Cross-Dataset Attack Success Rate of RRA.}}
    \resizebox{0.95\linewidth}{!}{
    \begin{tabular}{l|c|c|c|c}
    \hline
    \multicolumn{1}{l|}{\textbf{Dataset}} & \multicolumn{4}{c}{\textbf{Profiling Data}} \\
    \cline{2-5}
    Target Data & Synthetic~\cite{thedrcat2024llm}  & Medical~\cite{thedevastator2023medicalqa}  & Financial~\cite{saeedian2023financialqa}  & All Datasets \\
    \hline
    Synthetic     &  93.6\%  & 67.8\%  & 72.2\%   & 92.6\% \\
    Medical       & 79.3\%  &  91.5\% & 76.0\%  & 91.7\% \\
    Financial     &  86.3\% & 73.4\%  & 91.9\% & 91.4\% \\
    \hline
    \end{tabular}
    }
    \label{tab: rra-cross}
\end{table}

{\color{black}
\subsection{Robustness to System Noise}\label{sec: sensitivity study}
}
\update{
As the adversary exploits common resources shared with the victim for side-channel, there may be system noise on it when there are additional workloads using the resources concurrently. 
So far we focus on a common execution environment, where the system only has the operating system (OS) and its background processes running in addition to the foreground victim and spy processes. 
Note this execution environment is critical for us to establish the side-channels and investigate their working mechanisms and efficacy. 
Next we study the robustness of \mymethod under more realistic scenarios, where there are other independent users on the same machine simultaneously running the same victim MOE model, i.e., co-tenants using the LLM/VLM services. The robustness varies for the four proposed side-channels.}


\noindent\update{\textbf{Cache-occupancy channel.} The cache-occupancy channel leverages private L1/L2 caches of each physical core. When the co-tenant resides on other cores, which is typical under heavy workloads, the cache side-channel is immune to other co-tenants activities.}

\noindent\update{\textbf{GPU performance counter.} Since we use Nsight performance counters that expose the context ID of each launched kernel, we can reliably attribute observed GPU activity to specific processes. This allows us to distinguish victim kernels from those launched by other processes, even in multi-tenant environments. Therefore, our GPU performance counter side-channel remains robust and unaffected by concurrent processes, ensuring high fidelity of information leakage.}

\noindent\update{\textbf{Pageout+Reload and TLB Evict+Reload.} For these two side-channels, the adversary relies on monitoring the common MoE libraries that are loaded to shared memory pages or cached in the GPU TLB, respectively. These shared resources are susceptible to interference from co-tenants who are also using the MoE-based LLM/VLM, concurrently with the victim. Contention on the shared resources raises noise on the side-channel and may lead to degraded performance for information retrieval. Take the recovery of an expert sequence as an example, we vary the number of contending workloads and investigate the impact of such system noise on the recovery accuracy. 
Table \ref{tab:workload_impact} shows the experimental results, which indicate the accuracy declines as the resource competition increases. However, the decline diminishes when the competing workloads exceeds three.  
This is due to the hardware-limited degree of parallelism. 
Only up to three processes are scheduled for concurrent execution under the resource constraints. As a result, the level of interference saturates, and adding more competing processes does not introduce further noise into the side-channel observations.
}

\begin{table}[h]
  \centering
  \caption{Impact of Concurrent Workloads on Expert Sequence}
  \label{tab:workload_impact}
  \resizebox{0.9\linewidth}{!}{
  \begin{tabular}{@{}ccc@{}}
    \toprule
    \makecell[c]{\#\,Competing\\Workloads} &
    \textbf{Pageout+Reload (\%)} &
    \textbf{TLB~Evict+Reload (\%)} \\
    \midrule
      0 (System Noise) & 99.2 & 94.3 \\
      1                & 91.2 & 85.1 \\
      2                & 83.8 & 78.5 \\
      3                & 81.4 & 68.3 \\
      4                & 81.8 & 68.4 \\
    \bottomrule
  \end{tabular}
  }
\end{table}

{\color{black}
\subsection{Practicality Analysis of \mymethod} \label{sec: attack cost}
As detailed in Section~\ref{sec: end2end}, the \mymethod attack consists of two key stages: (1) offline profiling of the target MoE model's execution behavior, and (2) online side-channel analysis to capture information leakage. The practicality of \mymethod hinges on two critical aspects: the offline profiling cost and the stealthiness of the online SCA. This section analyzes these aspects.

\subsubsection{Profiling Cost}
The offline profiling phase of \mymethod is exclusively \textit{software-based}. It involves executing the target MoE model with specific inputs and only recording the resulting expert footprints (i.e., the expert load and expert sequence). 
It is platform-agnostic and no side-channel is involved during profiling. 
The primary cost is the computational overhead incurred by running the (substitute) victim model for collecting the expert footprints.

The profiling cost differs significantly between the two proposed LLM attacks (PIA and RRA) due to their distinct requirements:

\textbf{Prompt Inference Attack:} Profiling requires only the \textit{expert load} observed during the \textit{pre-filling phase} of a single inference run per prompt. 
Since the pre-filling phase is highly parallelizable and involves generating only the first output token, its execution time is short. 
As shown in Table~\ref{tab: pia cost}, generating the required profiling dataset (size: 10,000 prompts) is efficient, taking approximately 1 hour or less for the evaluated models.

\begin{table}[h]
    \centering
    \caption{PIA Profiling Cost for Different Models}
    \resizebox{0.95\linewidth}{!}{
    \begin{tabular}{l|c|c|c}
    \hline
    & DeepSeek-V2-Lite & Qwen-1.5b-MoE & TinyMixtral-MoE \\ 
    \hline
    Pre-filling Time per Prompt & 0.37s & 0.36s & 0.04s \\
    \hline
    PIA Profiling Cost &  1.02h  & 1.0h &  0.11h \\
    \hline
    \end{tabular}
    }
    \label{tab: pia cost}
\end{table}

\textbf{Response Reconstruction Attack:} In contrast, profiling for RRAs targets the \textit{decoding phase}, and requires capturing the \textit{expert sequence} for \textit{every output token} generated  for each prompt. 
This necessitates running the entire, sequential, decoding process for each prompt in the profiling dataset.  
Consequently, the profiling time scales with the total number of tokens generated across all profiling prompts. 
Table~\ref{tab: rra cost} presents the results for DeepSeek-V2-Lite across different datasets.
While the total profiling time is high, it is a one-time offline process for a selected target model. The subsequent next step of using the profiled expert sequence/load is training a prediction DNN model to map the footprints to prompts/tokens for attacks. Such model training is computationally inexpensive compared to the MoE model inference cost and the time can be neglected. 

\begin{table}[h]
    \centering
    \caption{RRA Profiling Cost (DeepSeek-V2-Lite)}
    \resizebox{1.0\linewidth}{!}{
    \begin{tabular}{l|c|c|c|c}
    \hline
    & Synthetic~\cite{thedrcat2024llm} & Medical~\cite{thedevastator2023medicalqa} & Financial~\cite{saeedian2023financialqa} & All \\ \hline
    Decoding time per token & 0.09s & 0.09s & 0.09s & 0.09s \\ \hline
    \# of tokens& 2978872 & 2561922 & 2158464 & 7699258 \\ \hline
    RRA profiling Time & 75h & 64h & 54h & 193h \\ \hline
    \end{tabular}
    }
    \label{tab: rra cost}
\end{table}

\subsubsection{Attack Stealthiness}
We assume that the adversary is co-located with the victim user on a physical machine. Collection of side-channel traces during victim inference would not interfere with the model execution much due to the dominant computation time for LLM inferences. The delay caused by side-channel measurements (e.g., due to eviction and reload at the hardware level) contributes only marginally to the overall latency.  For example, the cache occupancy channel incurs a 1.2\% increase in the computation time, the Pageout + Reload side-channel 4.4\%, the GPU performance counter side-channel 7.9\%, and the TLB Evict + Reload side-channel 5.0\%. With LLM/VLM services for multiple users, the victim MoE system is designed to support co-located users, meaning that certain inference delays are considered normal. Thus, our side-channel attacks cannot be detected easily by the victim from the model execution time, making them stealthy. 
}

\ad{Should we change the section title to Attack cost and stealthiness?}\rd{I prefer no change. I think the attack latency is one of the attack cost here--I think cost can better cover these two aspects.}

\section{Related Works} \label{sec: related works}

\subsection{Side-Channel Analysis} \label{bg: sca}
Side-channel attacks leverage indirect information obtained during system execution to extract sensitive data. Traditional SCA methods against DNNs leverage different types of leakages, such as power consumption~\cite{gao2024deeptheft}, electromagnetic emanation~\cite{yu2020deepem, gongye2024side}, and vulnerable architectural states (e.g., cache~\cite{yan2020cache, liu2024deepcache}, floating-point unit~\cite{gongye2020reverse}, and GPU context switching~\cite{wei2020leaky}), to recover the model architecture or parameters, an IP piracy. However, few studies have addressed the SCA vulnerabilities of model execution associated with user inputs, particularly for modern large language models.
In this work, \mymethod discovers a novel attack surface in MoE-based LLMs.
Beyond the established SCAs summarized in Table~\ref{tab: sca attak vector structure}, our \mymethod is the first of its kind for privacy evaluation via side-channel analysis, demonstrating that various architectural SCAs can compromise the privacy of MoE systems in multiple ways.

\subsection{Prompt Inference} \label{bg: llm attacks}
Inferring users’ prompts has attracted significant attention in LLM privacy research. Existing work can be categorized into two streams. The first focuses on software-based attacks~\cite{sha2024prompt, yang2024prsa}, which obtain input prompts and output responses from the system and use learning strategies to reconstruct users’ prompts.
In contrast, our proposed \mymethod complements these approaches by providing additional insights into LLM execution patterns of the MoE architecture.
The second stream addresses system-level security vulnerabilities in LLM services~\cite{wu2025know, song2024early}. For instance, Wu et al.\cite{wu2025know} exploit shared KV caches in multi-tenant LLM services to reconstruct users’ inputs by leveraging timing differences of controlled queries. Similarly, Song et al.\cite{song2024early} explore timing variations during accessing the shared KV caches to extract system prompts.
However, these vulnerabilities rely on outsourced, vulnerable model deployment services such as VLLM~\cite{kwon2023efficient} or SGLANG~\cite{zheng2312sglang}.
\mymethod specifically targets the MoE architecture of LLM models, which is becoming increasingly prevalent, opening up a new avenue for LLM privacy attacks.
\yf{what is input-output pair here?  is this profiling based?} \rd{revised, it is the input prompts and response  }
\yf{clarify the "outsourced model deployment services"} \rd{this is third party deployment services such as vllm and sglang.}
\yf{what is the cache timing variations? architectural cache or KV cache?} \rd{revised. KV cache.} 
\section{Discussion \& Conclusions} 
\subsection{Possible Mitigation}
To defend against \mymethod-based side-channel attacks, potential mitigation strategies can be categorized into three groups:

\noindent \textbf{Robust Training Algorithms.} One approach is to introduce randomness to the training process using optimization. For example, when training the gating layer of the MoE router, incorporating techniques such as differential privacy (DP)~\cite{dwork_differential_2006, abadi_deep_2016} can add random noise to the router decisions and limit information leakage.
However, integrating DP into the router may lead to imprecise expert selection, which could adversely affect the model performance. 
\update{Another possible way to mitigate \mymethod is introducing more general experts to the MoE structure. As shown in Section~\ref{sec: attack performance evaluation}, less-specialized experts have a smaller leakage compared to the fine-grained ones. However, such design may reduce the routing efficiency and affect model performance on specific sub-problems.}
\noindent \textbf{Secure Deployment Strategies.} Another line of defense involves considering the deployment environment during the design of the MoE model. Since \mymethod requires a sequential execution pattern for each expert, one possible mitigation is to randomize the execution order to obscure observing the expert patterns. 
This randomization, however, might introduce computational overhead. 
Alternatively, distributing the experts across multiple devices~\cite{deepseek-ai_deepseek-r1_2025, deepseek-deepep} could reduce architectural leakage; yet, this approach may still be vulnerable to side-channels during inter-device communication.

\noindent \textbf{Adaptation of General Side-Channel Mitigation.} Finally, defenses originally designed to counter general side-channel attacks can be adapted to protect against \mymethod-based attacks. 
For instance, balanced computation of LLM models~\cite{ahmadi2023dnn} can effectively obscure the expert load $\boldsymbol{L}$. More generally, restricting access to sensitive system tools such as performance counter(e,g, NVIDIA Nsight), system call(e.g, {\color{gray} \texttt{madvise()}}) and high-resolution timer can limit attackers’ ability to collect side-channel traces. Additionally, detecting anomalous background processes—e.g., those repeatedly accessing shared libraries or issuing high-frequency memory probes—can help identify \mymethod-style attacks in real time. 
\update{Rigorously isolating shared resources (e.g., cache, performance counters) could mitigate microarchitectural side-channel attacks at their source. Nevertheless, this mitigation may incur substantial performance overhead.}

\subsection{Future Work}
\update{
This work primarily investigates SCA vulnerabilities with single-device deployments of MoE models. 
To broaden the scope of our attacks, we outline two key directions for future research:}

\noindent\update{\textbf{Scalability to Distributed Deployments:} We plan to extend \mymethod to analyze large-scale MoE models deployed across heterogeneous hardware platforms. This includes investigating vulnerabilities in multi-GPU configurations~\cite{deepseek-deepep} and cross-CPU/GPU setups~\cite{ktransformers}. Such distributed environments introduce new potential side-channels, such as variations in inter-device communication latency and data transfer patterns, which may necessitate adaptations of our attack methodology and reveal novel privacy risks.}

\noindent\update{\textbf{Adaptation to Evolving MoE Architectures:} As MoE architectures rapidly evolve (e.g., DeepSeek-V3~\cite{deepseek-ai_deepseek_2024}, Llama 4~\cite{meta2025llama}), featuring diverse expert specialization strategies (e.g., mixtures of specialized and general experts), future work will focus on evaluating and adapting \mymethod to these newer designs. Understanding the security implications of these architectural innovations is crucial for ensuring robustness of future MoE-based LLM/VLM systems.
}

\subsection{Conclusions}
In this work, we introduce \mymethod, a novel attack surface that exposes privacy vulnerabilities in Mixture-of-Experts LLMs.
We identify four architectural side-channels that track both temporal and spatial footprints of expert module execution in MoE systems. Leveraging these side-channels, we develop four privacy attacks capable of recovering users’ inputs and system responses in language and multi-modal models. Our comprehensive evaluation demonstrates high attack effectiveness on CPU or GPU, underscoring the need to protect user privacy in the design of efficient inference algorithms for large-scale language and vision models.

\section{Ethical Concern}
\update{All models employed in this study are open-source, and all attack experiments were conducted exclusively on local laboratory servers with controlled access. Furthermore, the identified vulnerability has been reported to the open-source community (e.g., HuggingFace), and the microarchitectural side-channel vulnerabilities have been disclosed to the respective hardware manufacturers (e.g., NVIDIA). }

\bibliographystyle{ACM-Reference-Format}
\bibliography{sample-base}

\appendix
\newpage
\section{Implementation of DeepSeekMoE} \label{app: moe-implementation}
\begin{lstlisting}[
    caption={DeepSeek MoE Inference},
    label={lst:moe_infer_simplified}
]
func moe_infer(x, topk_ids, topk_weight):
    // 1. Count tokens for each expert
    tokens_per_expert = count_tokens(topk_ids)
    
    // 2. Sequentially call each expert
    outputs = []
    for expert_index in range(num_experts):
        num_tokens = tokens_per_expert[expert_index]
        if num_tokens == 0:
            continue
        
        // Get tokens for this expert
        expert_in = get_tokens(x, expert_index)
        // Call experts sequentially.
        expert_out = run_expert(expert_index, expert_in)
        outputs.append(expert_out)
        
    // 3. Aggregate outputs
    final_output = aggregate(outputs, topk_weight)
    return final_output
\end{lstlisting}

This is a simplified forward function for DeepSeekMoE~\cite{dai_deepseekmoe_2024}. The inputs to the function includes {\color{gray} \texttt{x}}--a tensor contains the input token(s), {\color{gray} \texttt{topk\_ids}}, which are the top-k indices of experts given each token (i.e., $\mathcal{I}$ in Section~\ref{sec: moe-description}), and {\color{gray} \texttt{topk\_weights}}, which are the weight to aggregate all experts for each token (i.e., $g_{i,t}$).
Function {\color{VioletRed} \texttt{count\_tokens}} counts the expert load for each experts and then the experts modules are execution \texttt{in a sequential manner} (the for loop). 
When we {\color{VioletRed} \texttt{get\_tokens}} for current expert, we will {\color{VioletRed} \texttt{run\_expert}} for all its input tokens.
If the model is deployed on a CPU, {\color{VioletRed} \texttt{run\_expert}} will be sequential, while for the single GPU-deployment, all tokens in  {\color{gray} \texttt{expert\_in}} will run in parallel.
In the end, the model will collect the output and {\color{VioletRed} \texttt{aggregate}} them as the real output.
Therefore, the key takeaway from the code snippet is that the experts are run one after another, and function {\color{VioletRed} \texttt{run\_expert}} may take various computational resources, leading to the SCA leakage of \mymethod.

\section{Diagram of Four Proposed Attacks} \label{app: attack diagrams}
\begin{figure}[h]
    \centering
    \includegraphics[width=\linewidth]{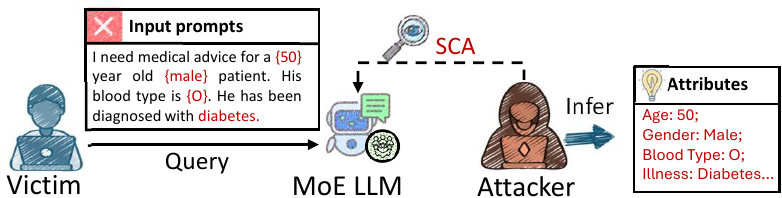}
    \caption{\texttt{Prompt Inference Attack}: the attacker aims to infer the user's input (i.e., private keywords).}
    \Description{Prompt Inference Attack}
    \label{fig: prompt-inference-attack}
\end{figure}

\begin{figure}[h]
    \centering
    \includegraphics[width=\linewidth]{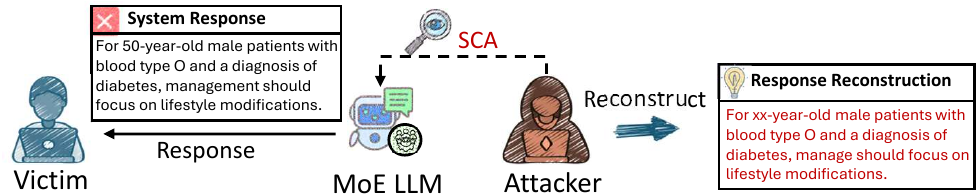}
    \caption{\texttt{Response Reconstruction Attack}: the attacker aims to recover the system response to the users.}
    \Description{Responsed Reconstruction Attack}
    \label{fig: response-reconstruction-attack}
\end{figure}

\begin{figure}[h]
    \centering
    \includegraphics[width=\linewidth]{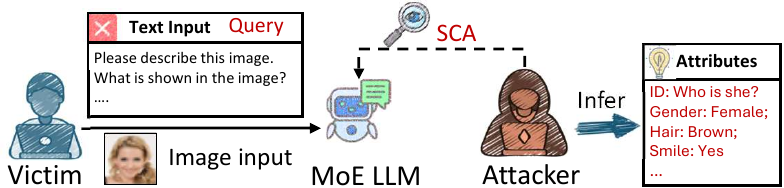}
    \caption{\texttt{Visual Inference Attack}: in a MOE-Vision-language model, the attacker aims to infer attributes of visual inputs.}
    \Description{Visual Attribute Inference Attack}
    \label{fig:attribute-inference-attack}
\end{figure}

\begin{figure}[h]
    \centering
    \includegraphics[width=\linewidth]{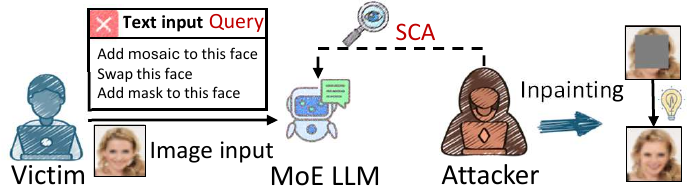}
    \caption{\texttt{Visual Reconstruction Attack}: the attacker aims to reconstruct the user's input with partial knowledge.}
    \Description{Visual Reconstruction Attack}
    \label{fig:inpainting-flow}
\end{figure}

\section{Model Structures Used in Attacks} \label{app: attack-structure}

As shown in Table~\ref{tab: appendix: attack model}, for PIA, RRA, and VIA, we adopt a simple classification model including MLP and Logistic Regression, which can already achieve a very high accuracy.
For VRA, we adopt a U-Net–style architecture that encodes the masker image via successive downsampling layers to a compact feature representation.
And then this latent vector is combined with a noise vector and condition features from \mymethod through a fully connected layer, refined by a self-attention mechanism at the bottleneck, and finally decoded with conditional upsampling blocks and skip connections to reconstruct a $32\times32$ inpainted image.

\begin{table}[h]
    \centering
    \caption{Model Structures used in Attacks}
    \resizebox{\linewidth}{!}{
    \begin{tabular}{l|c|l}
    \hline
    Attack & Model & Description \\
    \hline
    PIA and VIA  & $3$-Layer MLP &  \parbox{0.6\linewidth}{\vspace{1mm} The model used has the hidden size $[input, 1024,512,output]$ \vspace{1mm}}\\
    \hline
    RRA  & Logistic Regression & i.e., $1$-layer MLP \\
    \hline
    VRA  & Conditional Unet\cite{ronneberger2015u} & \parbox{0.6\linewidth}{\vspace{1mm} Unet with $4\times 4$ latent feature with the condition latent space size of $128$  \vspace{1mm}}\\ 
    \hline
    \end{tabular}
}
    \label{tab: appendix: attack model}
\end{table}

\section{Templates of Healthcare Datasets} \label{app: template}
The templates used in the PIA attack are shown in Table~\ref{tab: pia templates}.
\begin{table*}[h]
\scriptsize
    \centering
        \caption{Templates Used in PIA Attacks}
    \resizebox{\linewidth}{!}{
    \begin{tabular}{c|c}
    \hline
    Template     &  Length\\
    \hline
    \parbox{0.95\linewidth}{\vspace{1mm} "I need medical advice for a \textbf{[Age]} year old \textbf{[Gender]} patient. His blood type is \textbf{[Blood Type]}. He has been diagnosed with \textbf{[Illness]}. He is covered by \textbf{[Insurance]}, and the latest treatment bill was \textbf{[Billing Amount]}. What recommendations do you have for his condition and ongoing care?" \vspace{1mm}}     &  Short\\
    \hline
    \parbox{0.95\linewidth}{\vspace{1mm} "A \textbf{[Age]} year old  \textbf{[Gender]} patient with blood type \textbf{[Blood Type]} has been diagnosed with \textbf{[Illness]}. The patient is insured under \textbf{[Insurance]}, and the treatment cost was \textbf{[Billing Amount]}. What are the best management strategies for his condition?" \vspace{1mm}}     &  Short\\
    \hline
    \parbox{0.95\linewidth}{\vspace{1mm} "Can you provide medical guidance for a \textbf{[Age]} year old \textbf{[Gender]} patient? He has blood type \textbf{[Blood Type]} and has been diagnosed with \textbf{[Illness]}. His insurance is with \textbf{[Insurance]}, and the latest billing amount was \textbf{[Billing Amount]}. What treatment plans would be suitable?" \vspace{1mm}}     &  Short\\
    \hline
    \parbox{0.95\linewidth}{\vspace{1mm} "A \textbf{[Age]} year old \textbf{[Gender]} with blood type \textbf{[Blood Type]} is dealing with \textbf{[Illness]}. His insurance provider is \textbf{[Insurance]}, and the treatment expenses totaled \textbf{[Billing Amount]}. What care options should be considered?" \vspace{1mm}}     &  Short\\
    \hline
     \parbox{0.95\linewidth}{\vspace{1mm} "Seeking expert advice for a \textbf{[Age]} year old \textbf{[Gender]} with blood type \textbf{[Blood Type]}. Diagnosed with \textbf{[Illness]}, insured by  \textbf{[Insurance]}, and the latest medical expenses amounted to \textbf{[Billing Amount]}. How should this case be managed?" \vspace{1mm}}     &  Short\\ 
    \hline
    \parbox{0.95\linewidth}{\vspace{1mm} "I am consulting for a \textbf{[Age]} year old \textbf{[Gender]} patient whose blood type is \textbf{[Blood Type]}. He has been diagnosed with \textbf{[Illness]}, is insured by  \textbf{[Insurance]}, and his latest treatment costs were \textbf{[Billing Amount]}. What would be the optimal medical approach?" \vspace{1mm}}     &  Short\\ 
    \hline
    \parbox{0.95\linewidth}{\vspace{1mm}"A patient, \textbf{[Age]} years old and \textbf{[Gender]}, has blood type \textbf{[Blood Type]} and has recently been diagnosed with \textbf{[Illness]}. His insurance is with  \textbf{[Insurance]}, and his latest hospital bill was \textbf{[Billing Amount]}. What would be the recommended course of action?" \vspace{1mm}}     &  Short\\    
    \hline 
    \parbox{0.95\linewidth}{\vspace{1mm}"Medical consultation is needed for a \textbf{[Age]} year old \textbf{[Gender]} patient with blood type \textbf{[Blood Type]}. He is diagnosed with \textbf{[Illness]} and is covered by  \textbf{[Insurance]}. The last treatment charge was \textbf{[Billing Amount]}. What steps should be taken for effective care?" \vspace{1mm}}     &  Short\\ 
    \hline 
    \parbox{0.95\linewidth}{\vspace{1mm}"A \textbf{[Age]} year old \textbf{[Gender]} patient with \textbf{[Blood Type]} blood type has been diagnosed with \textbf{[Illness]}. He is insured through  \textbf{[Insurance]}, and the medical bill was \textbf{[Billing Amount]}. Could you guide on the best treatment options and follow-up care?" \vspace{1mm}}     &  Short\\    
    \hline 
    \parbox{0.95\linewidth}{\vspace{1mm}"I am looking for expert medical recommendations for a \textbf{[Age]} year old \textbf{[Gender]} patient. His blood type is \textbf{[Blood Type]}, and he has been diagnosed with \textbf{[Illness]}. He is insured under  \textbf{[Insurance]}, with a recent hospital bill of \textbf{[Billing Amount]}. What should be the next steps in his treatment plan?"\vspace{1mm}}     &  Short\\    
    \hline 
    \parbox{0.95\linewidth}{\vspace{1mm}"I am requesting comprehensive medical recommendations for a \textbf{[Age]}-year-old \textbf{[Gender]} patient who presents with blood type \textbf{[Blood Type]}. The patient was recently diagnosed with \textbf{[Illness]}, a condition that has begun to significantly impact his daily life and overall well-being. He is currently insured under the \textbf{[Insurance]} plan, which has provided partial coverage for treatments to date. The most recent treatment bill totaled \textbf{[Billing Amount]}, and we are evaluating both clinical and financial options moving forward. Given the complexity of the condition and the patient's age, we are seeking a balanced, evidence-based treatment approach that not only addresses immediate symptoms but also ensures sustainable long-term management. What therapies, interventions, or monitoring strategies would you recommend for this patient profile? Additionally, are there any recent guidelines or best practices that could inform our next steps? Please advise on both pharmacological and non-pharmacological options."\vspace{1mm}} & Long\\ \hline

\parbox{0.95\linewidth}{\vspace{1mm}"A \textbf{[Gender]} patient, aged \textbf{[Age]}, with blood type \textbf{[Blood Type]}, has been diagnosed with \textbf{[Illness]} and is now under active medical review. He is covered by \textbf{[Insurance]}, which has been helpful in managing the expenses, the most recent being \textbf{[Billing Amount]}. However, considering the progressive nature of the illness, we are seeking updated and detailed medical input. Our primary goal is to establish a tailored, multidisciplinary treatment plan that takes into account the patient's medical history, age, and overall health status. We're particularly interested in understanding how to structure care to reduce the risk of complications and improve prognosis. What diagnostic evaluations should be prioritized? Are there specific treatment algorithms or clinical pathways you would recommend based on current standards of care? Your insights into both acute and maintenance therapy strategies will be greatly appreciated."\vspace{1mm}} & Long\\ \hline

\parbox{0.95\linewidth}{\vspace{1mm}"We are evaluating the best possible treatment strategies for a \textbf{[Age]} year old \textbf{[Gender]} patient recently diagnosed with \textbf{[Illness]}. His blood type is \textbf{[Blood Type]}, and he is enrolled in a healthcare plan under \textbf{[Insurance]}. The most recent billing from treatment was \textbf{[Billing Amount]}, and while insurance has covered a portion, cost-effective yet evidence-based care is a priority. This condition presents unique challenges in patients of his demographic profile, and we are looking for expert recommendations on how to proceed. Could you provide guidance on current medical standards for managing \textbf{[Illness]} in a patient with these characteristics? We are open to both conventional and advanced treatment options, including lifestyle interventions, medication plans, and any recommended follow-up protocols. Furthermore, if there are any emerging therapies, trials, or alternative treatment approaches that have shown efficacy, we would be eager to consider them as well."\vspace{1mm}} & Long\\ \hline

\parbox{0.95\linewidth}{\vspace{1mm}"A \textbf{[Age]}-year-old \textbf{[Gender]} patient with blood type \textbf{[Blood Type]} has recently been diagnosed with \textbf{[Illness]}, and we are reviewing potential care strategies. The patient has health insurance through \textbf{[Insurance]}, and the latest billing from his treatment was \textbf{[Billing Amount]}. Given the complexity of this condition, we are aiming for a personalized, comprehensive management plan that goes beyond symptom control. What would be the most effective treatment modalities to consider for a patient with this profile? In particular, we are interested in guidance around early-stage interventions, lifestyle modifications, and long-term monitoring strategies. Could you also advise on how to prioritize treatment if financial resources become constrained despite partial insurance coverage? Additionally, what warning signs should we monitor that may necessitate urgent reevaluation or hospitalization? A thorough breakdown of both short-term and longitudinal care options would be highly valuable."\vspace{1mm}} & Long\\ \hline

\parbox{0.95\linewidth}{\vspace{1mm}"This inquiry concerns a \textbf{[Gender]} patient, aged \textbf{[Age]}, with a confirmed diagnosis of \textbf{[Illness]} and blood type \textbf{[Blood Type]}. The patient is insured via \textbf{[Insurance]}, with the latest treatment bill totaling \textbf{[Billing Amount]}. Our clinical team is seeking expert advice to guide the next phase of medical management. The patient has shown both expected and unexpected responses to initial treatment, and we now face several options for continued care. We would appreciate an expert opinion on which therapeutic pathways have the highest success rates in similar patient populations. Are there any advanced diagnostic tools, imaging, or biomarkers that could provide further clarity? Additionally, how should we structure follow-ups, and what supportive care measures should be considered? We also welcome recommendations regarding patient education, lifestyle adaptation, and psychosocial support, as these are integral to our holistic care strategy."\vspace{1mm}} & Long\\ \hline

\parbox{0.95\linewidth}{\vspace{1mm}"Could you offer detailed medical advice for a \textbf{[Age]}-year-old \textbf{[Gender]} patient who has recently been diagnosed with \textbf{[Illness]}? His blood type is \textbf{[Blood Type]}, and he is currently insured through \textbf{[Insurance]}. The last treatment bill amounted to \textbf{[Billing Amount]}. The patient’s condition is currently stable but may progress without timely and targeted intervention. Our goal is to ensure an evidence-based care plan is implemented that will improve his quality of life and minimize risks over time. Given the patient’s age and the nature of \textbf{[Illness]}, what treatment modalities, both medical and behavioral, would be appropriate? Additionally, could you provide insights into how comorbidities or risk factors may alter the recommended treatment course? Please include any relevant updates to clinical guidelines or innovative approaches being used in similar cases."\vspace{1mm}} & Long\\ \hline

\parbox{0.95\linewidth}{\vspace{1mm}"We are managing care for a \textbf{[Age]} year old \textbf{[Gender]} patient with blood type \textbf{[Blood Type]}, who has been diagnosed with \textbf{[Illness]}. The patient is insured by \textbf{[Insurance]}, which has partially reimbursed a recent medical expense of \textbf{[Billing Amount]}. In light of the diagnosis, we are carefully planning a management protocol that takes into account both clinical efficacy and cost-efficiency. Given your expertise, could you share recommendations regarding the optimal sequence of treatments for this condition? Should we consider combination therapies, and are there particular drugs or devices that have shown promise in similar patient cohorts? Furthermore, are there specific clinical milestones we should monitor during the treatment process to evaluate effectiveness or guide adjustment? Your input will be used to refine a personalized and sustainable treatment plan for this individual."\vspace{1mm}} & Long\\ \hline

\parbox{0.95\linewidth}{\vspace{1mm}"A \textbf{[Gender]} patient, age \textbf{[Age]}, has been diagnosed with \textbf{[Illness]} and has blood type \textbf{[Blood Type]}. The individual is currently covered by \textbf{[Insurance]}, and their latest treatment expenses totaled \textbf{[Billing Amount]}. The patient has begun exhibiting symptoms consistent with the condition’s progression, and we are now in the process of reviewing updated treatment protocols. Our team would appreciate expert insight into current standards of care, particularly for patients in this age group. We are especially interested in strategies that balance efficacy, accessibility, and long-term health outcomes. What baseline evaluations should be revisited at this stage? Which medications or therapeutic techniques have demonstrated consistent success, and are there non-pharmacological alternatives worth considering in parallel? Additionally, are there psychosocial or lifestyle considerations that may impact adherence and outcomes?"\vspace{1mm}} & Long\\ \hline

\parbox{0.95\linewidth}{\vspace{1mm}"This case involves a \textbf{[Age]}-year-old \textbf{[Gender]} patient who has been diagnosed with \textbf{[Illness]}. His blood type is \textbf{[Blood Type]}, and he is covered under \textbf{[Insurance]}. The most recent medical invoice came to \textbf{[Billing Amount]}, and our care team is looking for advice on designing the most effective treatment roadmap moving forward. Given the patient’s demographics and diagnosis, what would constitute the best practices in current medical literature? Could you advise on any known prognostic indicators, early warning signs, or escalation triggers we should be monitoring? We are particularly interested in integrative strategies that involve both pharmaceutical and lifestyle interventions. Also, if there are any screening tools or checklists you recommend for ensuring comprehensive care, we’d greatly appreciate access to such frameworks."\vspace{1mm}} & Long\\ \hline

\parbox{0.95\linewidth}{\vspace{1mm}"I am reaching out on behalf of a \textbf{[Age]}-year-old \textbf{[Gender]} patient diagnosed with \textbf{[Illness]}, whose blood type is \textbf{[Blood Type]}. He is currently under a healthcare plan from \textbf{[Insurance]}, and the most recent treatment invoice was \textbf{[Billing Amount]}. While initial evaluations and stabilization were successful, we are now developing a full-scale treatment plan for ongoing care. What guidelines or evidence-based practices should inform our decisions at this stage? Specifically, we are looking for advice on medication choice, monitoring frequency, and any supportive therapies that should be concurrently deployed. Furthermore, we seek insights into patient-centered care models that could improve adherence and outcomes, especially in patients of this age and health profile. If any peer-reviewed case studies or clinical trials have explored similar cases, we’d be very interested in applying those learnings."\vspace{1mm}} & Long\\ \hline

    \end{tabular}

    }
    \label{tab: pia templates}
\end{table*}

\end{document}